\documentclass[twocolumn,10pt]{IEEEtran}
\usepackage{ifpdf, flushend}

\ifCLASSINFOpdf
  \usepackage[pdftex]{graphicx}
  \graphicspath{{../pdf/}{../jpeg/}}
  \DeclareGraphicsExtensions{.pdf,.jpeg,.png}
\else
  \usepackage[dvips]{graphicx}
  \graphicspath{{../eps/}}
  \DeclareGraphicsExtensions{.eps}
\fi
\usepackage[cmex10]{amsmath}
\usepackage {amssymb}
\usepackage{algorithmic}
\usepackage{array}
\usepackage{mdwmath}
\usepackage{mdwtab}
\usepackage{eqparbox}
\usepackage{url}
\usepackage{hyperref}
\usepackage{algorithm}
\usepackage{algorithmic}

\newcommand{\argmax}{\operatornamewithlimits{argmax}}
\newcommand{\beq}{\begin{equation}}
\newcommand{\eeq}{\end{equation}}
\newcommand{\beqn}{\begin{eqnarray}}
\newcommand{\eeqn}{\end{eqnarray}}
\newcommand{\beqno}{\begin{eqnarray*}}
\newcommand{\eeqno}{\end{eqnarray*}}
\newcommand{\bma}{\begin{displaymath}}
\newcommand{\ema}{\end{displaymath}}
\newcommand{\bnu}{\begin{enumerate}}
\newcommand{\enu}{\end{enumerate}}
\newcommand{\bce}{\begin{center}}
\newcommand{\ece}{\end{center}}
\newcommand{\btb}{\begin{tabular}}
\newcommand{\etb}{\end{tabular}}

\begin{document}

%
\title{Design and Optimal Configuration of Full-Duplex MAC Protocol for Cognitive Radio Networks
Considering Self-Interference}

\author{Le~Thanh~Tan,~\IEEEmembership{Member,~IEEE,} Long~Bao~Le,~\IEEEmembership{Senior Member,~IEEE} 
\thanks{ Manuscript received November 08, 2015; accepted December 10, 2015. 
The editor coordinating the review of this paper and approving it for publication is  Dr. Wei Wang. 
}
\thanks{The  authors  are  with  the  Institut  National  de  la  Recherche  Scientifique--
\'{E}nergie, Mat\'{e}riaux et T\'{e}l\'{e}communications, Universit\'{e} du Québec, Montr\'{e}al,
QC J3X 1S2, Canada (e-mail: lethanh@emt.inrs.ca; long.le@emt.inrs.ca)}
}



\markboth{IEEE Access} {Le \MakeLowercase{\textit{et al.}}: Design and Optimal Configuration of Full-Duplex MAC Protocol for Cognitive Radio Networks Considering Self-Interference}

\maketitle

\begin{abstract}
\boldmath
In this paper, we propose an adaptive Medium Access Control (MAC) protocol for full-duplex (FD) cognitive radio networks in which
FD secondary users (SUs) perform channel contention followed by concurrent spectrum sensing and transmission,
and transmission only with maximum power in two different stages (called the FD sensing and transmission stages, respectively)
in each contention and access cycle. The proposed FD cognitive MAC (FDC-MAC) protocol does not require synchronization among SUs
and it efficiently utilizes the spectrum and mitigates the self-interference in the FD transceiver.
We then develop a mathematical model to analyze the throughput performance of the FDC-MAC protocol where 
both half-duplex (HD) transmission (HDTx) and FD transmission (FDTx) modes are considered in the transmission stage.
Then, we study the FDC-MAC configuration optimization through adaptively
controlling the spectrum sensing duration and transmit power level in the FD sensing stage
where we prove that there exists optimal sensing time and transmit power to achieve the maximum throughput
and we develop an algorithm to configure the proposed FDC-MAC protocol. Extensive numerical results are presented 
to illustrate the characteristic of the optimal FDC-MAC configuration and the impacts of protocol
parameters and the self-interference cancellation quality on the throughput performance.
Moreover, we demonstrate the significant throughput gains of the FDC-MAC protocol 
with respect to existing half-duplex MAC (HD MAC) and single-stage FD MAC protocols.

\end{abstract}

\IEEEpeerreviewmaketitle

\section{Introduction}

Engineering MAC protocols for efficient sharing of white spaces is an important research topic in cognitive radio networks (CRNs).
One critical requirement for the cognitive MAC design is that transmissions on the licensed frequency band from primary users (PUs) should  be satisfactorily protected from the SUs' spectrum
 access. Therefore, a cognitive MAC protocol for the secondary network must realize both the spectrum sensing and access functions so that timely detection of the PUs' communications and effective 
spectrum sharing among SUs can be achieved. Most existing research works on cognitive MAC protocols have focused on the design and analysis of HD MAC 
(e.g., see \cite{Cor09}--\cite{Zou12} and the references therein). 

Due to the HD constraint, SUs typically employ a two-stage sensing/access procedure where they perform spectrum sensing in the first stage before accessing
available spectrum for data transmission in the second stage \cite{Liang08}--\cite{wang08}. This constraint also requires SUs be synchronized during
the spectrum sensing stage, which could be difficult to achieve in practice.   
In fact, spectrum sensing enables SUs to detect white spaces that are not occupied by PUs \cite{Yu09}--\cite{Tan12}, \cite{Haykin09, Axell12}; 
therefore, imperfect spectrum sensing can reduce the spectrum utilization due to  failure in detecting white spaces and potentially result in collisions 
with active PUs. Consequently, sophisticated design and parameter configuration of cognitive MAC protocols must be conducted to
achieve good performance while appropriately protecting PUs \cite{Cor09}, \cite{Tan11}--\cite{wang08}, \cite{Konda08}.  
As a result, traditional MAC protocols \cite{Thorpe14}--\cite{Akyildiz99} 
adapted to the CRN may not provide satisfactory performance.

In general, HD MAC protocols may not exploit white spaces very efficiently since significant sensing time may be required, which would otherwise be utilized for data transmission. 
Moreover, SUs may not timely detect the PUs' activity during their transmissions, which can cause severe interference to active PUs.
Thanks to recent advances on FD technologies, a FD radio can transmit and receive data simultaneously on the same frequency band \cite{Duarte12}--\cite{Jain11}. 
One of the most critical issues of wireless FD communication is the presence of self-interference, which is caused by power leakage from the transmitter to the receiver of a FD transceiver.
The self-interference may indeed lead to serious communication performance degradation of FD wireless systems.
Despite recent advances on self-interference cancellation (SIC) techniques \cite{Everett14}--\cite{Korpi14} 
(e.g., propagation SIC, analog-circuit SIC, and digital baseband SIC), 
self-interference still exists due to various reasons such as the limitation of hardware and channel estimation errors.

\subsection{Related Works}

There are some recent works that propose to exploit the FD communications for MAC-level channel access in multi-user wireless networks
\cite{Jain11}--\cite{Choi15}. 
In \cite{Jain11}, the authors develop a centralized MAC protocol to support asymmetric data traffic where network nodes may transmit data 
packets of different lengths, and they propose to mitigate the hidden node problem by employing a busy tone.
To overcome this hidden node problem, Duarte et al. propose to adapt the standard 802.11 MAC protocol with the RTS/CTS handshake in \cite{Duarte14}.
Moreover, Goyal et al. in \cite{Goyal13} extend this study to consider interference between two nodes due to their concurrent transmissions.
Different from conventional wireless networks, designing MAC protocols in CRNs is more challenging because the spectrum sensing function must be
efficiently integrated into the MAC protocol.
In addition, the self-interference must be carefully addressed in the simultaneous spectrum sensing and access to mitigate its negative
impacts on the sensing and throughput performance.

The FD technology has been employed for more efficient spectrum access design in cognitive radio networks
 \cite{Cheng14}--\cite{Kim15}  where SUs can perform sensing and transmission simultaneously. 
In \cite{Cheng14}, a FD MAC protocol is developed which allows simultaneous spectrum access of the SU and PU networks 
where both PUs and SUs are assumed to employ the $p$-persistent MAC protocol for channel contention resolution and access.
This design is, therefore, not applicable to the hierarchical spectrum access in the CRNs where PUs should have higher spectrum access priority 
compared to SUs. 

In our previous work \cite{report}, we propose the FD MAC protocol by using the standard backoff mechanism as
in the 802.11 MAC protocol where we employ concurrent FD sensing and access during data transmission as well as frame fragmentation.
Moreover, engineering of a cognitive FD relaying network is 
considered in \cite{Kim12, Kim15}, where various resource allocation algorithms to improve the outage probability are proposed.
In addition, the authors in \cite{Ramirez13} develop the joint routing and distributed resource allocation for FD wireless networks.
In \cite{Choi15}, Choi et al. study the distributed power allocation for a hybrid FD/HD system where all network nodes operate in the HD mode but 
the access point (AP) communicates by using the FD mode.
In practice, it would be desirable to design an adaptable MAC protocol, which can be configured to operate in an
optimal fashion depending on specific channel and network conditions. This design will be pursued in our current work. 


\subsection{Our Contributions}

In this paper, we make a further bold step in designing, analyzing, and optimizing an adaptive
 FDC--MAC protocol for CRNs, where the self-interference and imperfect spectrum
sensing are explicitly considered. In particular, the contributions of this paper can be summarized 
as follows.

\begin{enumerate}

\item We propose a novel FDC--MAC protocol that can efficiently exploit the FD transceiver for 
spectrum spectrum sensing and access of the white space without requiring synchronization among SUs. 
In this protocol, after the $p$-persistent based channel contention phase, the winning SU enters
the data phase consisting of two stages, i.e., concurrent sensing and transmission
in the first stage (called FD sensing stage) and transmission only in the second stage (called transmission stage).
The developed FDC--MAC protocol, therefore, enables the optimized configuration of transmit
power level and sensing time during the FD sensing stage to mitigate the self-interference and appropriately protect the active PU.
After the FD sensing stage, the SU can transmit with the maximum power to achieve the highest throughput.  

\item We develop a mathematical model for throughput performance analysis of the proposed FDC-MAC protocol 
considering the imperfect sensing, self-interference effects, and the dynamic status changes of the PU.
In addition, both one-way and two-way transmission scenarios, which are called  HD transmission
(HDTx) and FD transmission (FDTx) modes, respectively, are considered in the analysis. Since the PU can change its idle/active 
status during the FD sensing and transmission stages, different potential status-change scenarios are studied in the analytical model.

\item We study the optimal configuration of FDC-MAC protocol parameters including the
 SU's sensing duration and transmit power to maximize the achievable throughput under
both FDTx and HDTx modes. We prove that there exists an optimal sensing time to
achieve the maximum throughput for a given transmit power value during the FD sensing
stage under both FDTx and HDTx modes. Therefore, optimal protocol
parameters can be determined through standard numerical search methods.

\item
Extensive numerical results are presented to illustrate the impacts of different 
protocol parameters on the throughput performance and the optimal configurations of the proposed FDC-MAC protocol.
Moreover, we show the significant throughput enhancement of the proposed FDC-MAC protocol compared to existing cognitive
MAC protocols, namely the HD MAC protocol and a single-stage FD MAC protocol with concurrent sensing and access during the whole data phase.
Specifically, our FDC-MAC protocol achieves higher throughput with the increasing maximum power while
the throughput of the single-stage FD MAC protocol decreases with the maximum power in the high power regime
due to the self-interference. Moreover, the proposed FDC-MAC protocol significantly outperforms the HD MAC protocol in terms of
system throughput.

\end{enumerate}

The remaining of this paper is organized as follows. Section ~\ref{SystemModel} describes the system and PU models. FDC--MAC protocol design, and
throughput analysis for the proposed FDC--MAC protocol are performed in Section ~\ref{MACNonFrag}.
Then, Section ~\ref{FDC_MAC_Configuration} studies the optimal configuration of the proposed FDC--MAC protocol to achieve the
maximum secondary throughput.
Section ~\ref{Results} demonstrates numerical results followed by concluding remarks in Section ~\ref{conclusion}.

\section{System and PU Activity Models}
\label{SystemModel}

\subsection{System Model}
\label{System}

We consider a cognitive radio network where $n_0$ pairs of SUs opportunistically exploit white spaces on 
one channel for communications. We assume that each SU is equipped with a FD transceiver; hence, the SUs can perform sensing and transmission simultaneously.
However, the sensing performance of each SU is affected by the self-interference from its transmitter since the transmitted power is leaked into
the received signal. We denote $I(P)$ as the average self-interference power, which is modeled as
 $I(P) = \zeta \left(P\right)^{\xi}$ \cite{Duarte12} where $P$ is the SU's transmit power, $\zeta$ and $\xi$ ($0 \leq \xi \leq 1$) are 
predetermined coefficients which represent the quality of self-interference cancellation (QSIC).
In this work, we design a asynchronous cognitive MAC protocol where no synchronization is required among SUs and between SUs and the PU.
We assume that different pairs of SUs can overhear transmissions from the others (i.e., a collocated network is assumed). 
In the following, we refer to pair $i$ of SUs as SU $i$ for brevity.

\subsection{Primary User Activity}
\label{PUAM}

We assume that the PU's idle/active status follows two independent random processes. We say that the channel is available and busy 
for SUs' access if the PU is in the idle and active (or busy) states, respectively. Let $\mathcal{H}_0$ and $\mathcal{H}_1$ denote the events 
that the PU is idle and active, respectively. To protect the PU, we assume that SUs must stop their transmissions and evacuate from the busy channel 
within the maximum delay of $T_{\sf eva}$, which is referred to as channel evacuation time. 

Let $\tau_{\sf ac}$ and $\tau_{\sf id}$ denote the random variables which represent the durations of active and idle channel states, respectively.
We denote probability density functions (pdf) of $\tau_{\sf ac}$ and $\tau_{\sf id}$ as $f_{\tau_{\sf ac}}\left(t\right)$ and $f_{\tau_{\sf id}}\left(t\right)$, respectively. 
While most results in this paper can be applied to general pdfs $f_{\tau_{\sf ac}}\left(t\right)$ and $f_{\tau_{\sf id}}\left(t\right)$, we mostly consider the
exponential pdf in the analysis.
In addition, let  $\mathcal{P}\left(\mathcal{H}_0\right) = \frac{{\bar \tau}_{\sf id}}{{\bar \tau}_{\sf id}+{\bar \tau}_{\sf ac}}$ and  $\mathcal{P}\left( \mathcal{H}_1 \right) = 
1 - \mathcal{P}\left(\mathcal{H}_0\right)$ present the probabilities that the channel is available and busy, respectively
where ${\bar \tau}_{\sf id}$ and ${\bar \tau}_{\sf ac}$ denote the average values of $\tau_{\sf ac}$ and $\tau_{\sf id}$, respectively.
We assume that the probabilities that $\tau_{\sf ac}$ and $\tau_{\sf id}$ are smaller than $T_{\sf eva}$ are sufficiently small (i.e., the PU
changes its status slowly) so that we can ignore events with multiple idle/active status changes in one channel evacuation interval $T_{\sf eva}$.

\section{Full-Duplex Cognitive MAC Protocol}
\label{MACNonFrag}

In this section, we describe the proposed FDC-MAC protocol and conduct its throughput
analysis considering imperfect sensing, self-interference of the FD transceiver, and
dynamic status change of the PUs.

\subsection{FDC-MAC Protocol Design}

\begin{figure*}[!t]
\centering
\includegraphics[width=150mm]{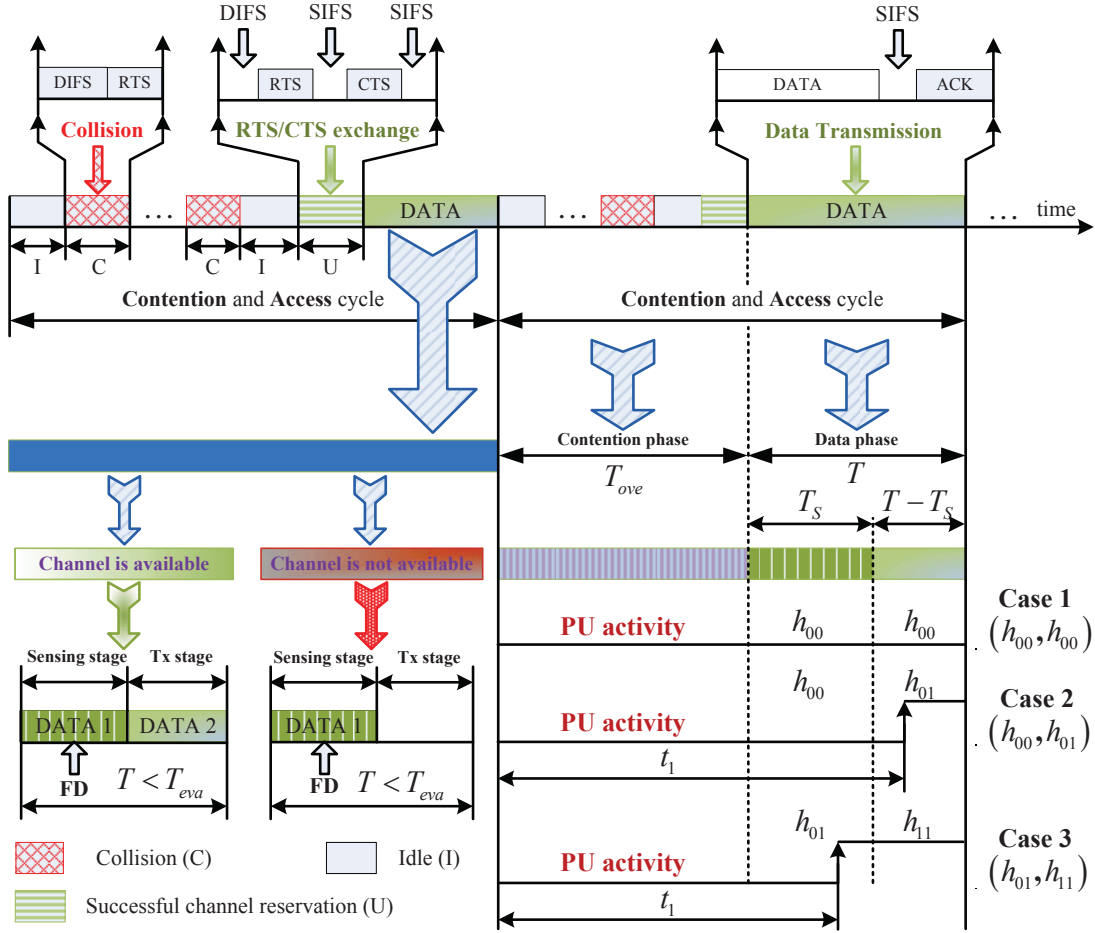}
\caption{Timing diagram of the proposed full-duplex cognitive MAC protocol.}
\label{Sentime_FDMAC_SF}
\end{figure*}

The proposed FDC-MAC protocol integrates three important elements of a cognitive MAC protocol,
namely contention resolution, spectrum sensing, and access functions.
Specifically, SUs employ the $p$-persistent CSMA principle \cite{Cali00} for contention resolution
where each SU with data to transmit attempts to capture an available channel with a probability $p$ after the channel is sensed to be idle during the standard DIFS interval  (DCF Interframe Space). 
If a particular SU decides not to transmit (with probability of $1-p$), it will sense the channel and attempt to transmit again in the next slot of length $\sigma$ 
with probability $p$.
To complete the reservation, the four-way handshake with Request-to-Send/Clear-to-Send (RTS/CST) exchanges \cite{Bian00} is employed to reserve 
the available channel for transmission.
Specifically, the secondary transmitter sends RTS to the secondary receiver and waits until it successfully receives the CTS from the secondary receiver. 
All other SUs, which hear the RTS and CTS exchange from the winning SU, defer to access the channel for a duration equal to the data transmission time, $T$.
Then, an acknowledgment (ACK) from the SU's receiver is transmitted to its corresponding transmitter to notify the successful reception of a packet.
Furthermore, the standard small interval, namely SIFS (Short Interframe Space), is used before the transmissions of CTS, ACK, and data frame as in the
standard 802.11 MAC protocol \cite{Bian00}.

In our design, the data phase after the channel contention phase comprises two stages where the SU performs
concurrent sensing and transmission in the first stage with duration $T_S$ and transmission only in the second stage with duration $T-T_S$.
Here, the SU exploits the FD capability of its transceiver to realize concurrent sensing and transmission 
the first stage (called FD sensing stage)  where the sensing outcome at the end of this stage (i.e., an idle or busy channel status) determines its 
further actions as follows.
Specifically, if the sensing outcome indicates an available channel then the SU transmits data in the second stage; otherwise, it remains silent for the remaining time of the 
data phase with duration $T-T_S$.

We assume that the duration of the SU's data phase $T$ is smaller than the channel evacuation time $T_{\sf eva}$ so timely evacuation from the busy channel can be 
realized with reliable FD spectrum sensing. Therefore, our design allows to protect the PU with evacuation delay at most $T$ if
the MAC carrier sensing during the contention phase
and the FD spectrum sensing in the data phase are perfect. Furthermore, we assume that the SU transmits at power levels $P_{\sf sen} \leq P_{\sf max}$ and 
$P_{\sf dat} = P_{\sf max}$ during
the FD sensing and transmission stages, respectively where $P_{\sf max}$ denotes the maximum power and
 the transmit power $P_{\sf sen}$ in the FD sensing stage will be optimized to effectively mitigate the 
self-interference and achieve good sensing-throughput tradeoff. The timing diagram of the proposed FDC--MAC protocol is illustrated 
in Fig.~\ref{Sentime_FDMAC_SF}.

We allow two possible operation modes in the transmission stage. 
The first is the HD transmission mode (HDTx mode) where there is only one direction of data transmission from the SU transmitter to the SU receiver.
In this mode, there is no self-interference in the transmission stage. The second is the FD transmission mode (FDTx mode) where two-way communications
between the pair of SUs are assumed (i.e., there are two data flows between the two SU nodes in opposite directions).
In this mode, the achieved throughput can be potentially enhanced (at most doubling the throughput of the HDTx mode) but self-interference must be 
taken into account in throughput quantification.

Our proposed FDC--MAC protocol design indeed enables flexible and adaptive configuration, which can efficiently exploit the capability of the FD transceiver.  
Specifically, if the duration of the FD sensing stage is set equal to the duration of the whole data phase (i.e., $T_S = T$), then the SU performs 
concurrent sensing and transmission for the 
whole data phase as in our previous design \cite{report}. This configuration may degrade the achievable throughput since the transmit power during the 
FD sensing stage is typically set smaller  $P_{\sf max}$  to mitigate the self-interference and achieve the required sensing performance. We will refer
the corresponding MAC protocol with $T_S = T$ as one-stage FD MAC in the sequel.

Moreover, if we set the SU transmit power $P_{\sf sen}$ in the sensing stage equal to zero, i.e., $P_{\sf sen} = 0$, 
then we achieve the traditional two-stage cognitive HD  MAC protocol where sensing and transmission are performed sequentially in two
 different stages \cite{Tan11, Tan12}. Moreover, the proposed FDC--MAC protocol is more flexible than existing designs \cite{report}, \cite{Tan11, Tan12} 
since different existing designs can be achieved through suitable configuration of its protocol parameters. 
It will be demonstrated later that the proposed FDC--MAC protocol achieves significant better throughput than that of the existing cognitive MAC protocols.
In the following, we present the throughput analysis based on which the protocol configuration optimization can be performed.

\subsection{Throughput Analysis}
\label{Tput_Ana_MACNonFrag}

We now conduct the saturation throughput analysis for the secondary network where all SUs are assumed to always have data to transmit.
The resulting throughput can be served as an upper bound for the throughput in the non-saturated scenario \cite{Bian00}.
This analysis is performed by studying one specific contention and access cycle (CA cycle) with the contention phase and data phase
 as shown in Fig.~\ref{Sentime_FDMAC_SF}.
Without loss of generality, we will consider the normalized throughput achieved per one unit  of system bandwidth (in bits/s/Hz). 
Specifically, the normalized throughput of the FDC--MAC protocol can be expressed as 
\beqn
\label{NT_NonFrag}
\mathcal{NT} = \frac{\mathcal{B}}{T_{\sf ove} + T},
\eeqn
where $T_{\sf ove}$ represents the time overhead required for one successful channel reservation (i.e., successful RTS/CTS exchanges), $T$
denotes the packet transmission time, and  $\mathcal{B}$ denotes the amount of data (bits) transmitted in one CA cycle per one unit of system bandwidth, which is expressed in bits/Hz.
To complete the throughput analysis, we derive the quantities $T_{\sf ove}$ and $\mathcal{B}$ in the remaining of this subsection.

\subsubsection{Derivation of $T_{\sf ove}$}

The average time overhead for one successful channel reservation can be calculated as
\beqn \label{tover}
T_{\sf ove} = {\overline T}_{\sf cont} + 2SIFS + 2PD + ACK,
\eeqn
where $ACK$ is the length of an ACK message, $SIFS$ is the length of a short interframe space, and $PD$ is the propagation delay
where $PD$ is usually small compared to the slot size $\sigma$, and ${\overline T}_{\sf cont}$ denotes the average time overhead
due to idle periods, collisions, and successful transmissions of RTS/CTS messages in one CA cycle. For
better presentation of the paper, the derivation of ${\overline T}_{\sf cont}$ is given in Appendix \ref{AppenA0}.

\subsubsection{Derivation of $\mathcal{B}$}

To calculate $\mathcal{B}$, we consider all possible cases that capture the activities of SUs and status changes of the PU in the FDC-MAC data phase of duration $T$.
Because the PU's activity is not synchronized with the SU's transmission, the PU can change its idle/active status any time.
We assume that there can be at most one transition between the idle and active states of the PU during one data phase interval. 
This is consistent with the assumption on the slow status changes of the PU as described in Section~\ref{PUAM} since $T < T_{\sf eva}$. 
Furthermore, we assume that the carrier sensing of the FDC-MAC protocol is perfect; therefore, the PU is idle at the beginning of the FDC-MAC data phase.
Note that the PU may change its status during the SU's FD sensing or transmission stage, which requires us to consider different possible events in the data phase. 

We use $h_{ij}$ ($i, j \in \left\{ 0, 1 \right\}$) to represent events capturing status changes of the PU in the FD sensing stage and transmission stage
where $i$ = 0 and $i$ = 1 represent the idle and active states of the PU, respectively.
For example, if the PU is idle during the FD sensing stage and becomes active during the transmission stage, then we represent
this event as $\left(h_{00}, h_{01}\right)$ where sub-events $h_{00}$ and $h_{01}$ represent the status changes in the FD sensing
and transmission stages, respectively. Moreover, if the PU changes from the idle to the active state during
the FD sensing stage and remains active in the remaining of the data phase, then we represent this event as $\left(h_{01}, h_{11}\right)$

It can be verified that we must consider the following three cases with the corresponding status changes of the PU during the FDC-MAC data phase to analyze $\mathcal{B}$.
\begin{itemize}
\item \textbf{Case 1}: The PU is idle for the whole FDC-MAC data phase  (i.e., there is no PU's signal in both FD sensing and transmission stages) and we denote this
event as $\left(h_{00}, h_{00}\right)$. The average number of bits (in bits/Hz) transmitted during the data phase in this case is denoted as $\mathcal{B}_1$.

\item \textbf{Case 2}: The PU is idle during the FD sensing stage but the PU changes from the idle to the active status in the transmission stage. 
We denote the event corresponding to this case as $\left(h_{00}, h_{01}\right)$ where $h_{00}$ and $h_{01}$ capture the sub-events in the FD sensing and transmission stages, respectively.
The average number of bits (in bits/Hz) transmitted during the data phase in this case is represented by $\mathcal{B}_2$.

\item \textbf{Case 3}: The PU is first idle then becomes active during the FD sensing stage and it remains active during the whole transmission stage.
Similarly we denote this event as $\left(h_{01}, h_{11}\right)$ and the average number of bits (in bits/Hz) transmitted during the data phase in this case is denoted as $\mathcal{B}_3$.
\end{itemize}

Then, we can calculate $\mathcal{B}$ as follows:
\beqn
\mathcal{B} = \mathcal{B}_1 + \mathcal{B}_2 + \mathcal{B}_3.
\eeqn
To complete the analysis, we will need to derive $\mathcal{B}_1$, $\mathcal{B}_2$, and $\mathcal{B}_3$, which are given in Appendix \ref{AppenB1}.

\section{FDC--MAC Protocol Configuration for Throughput Maximization}
\label{FDC_MAC_Configuration}

In this section, we  study the optimal configuration of the proposed FDC--MAC protocol to achieve the maximum 
 throughput while satisfactorily protecting the PU. 

\subsection{Problem Formulation}
\label{TputOpt}

Let $\mathcal{NT}(T_S, p, P_{\sf sen})$ denote the normalized secondary throughput, which is the function of the sensing time $T_S$, transmission probability
 $p$, and the SU's transmit power $P_{\sf sen}$ in the FD sensing stage. 
In the following, we assume a fixed frame length $T$, which is set smaller the required evacuation time $T_{\sf eva}$ to achieve timely evacuation from a busy channel for
the SUs. We are interested in determining suitable configuration for $p$, $T_S$ and $P_{\sf sen}$ to maximize the secondary throughput, $\mathcal{NT}(T_S, p, P_{\sf sen})$.
In general, the optimal  transmission probability $p$ should balance between reducing collisions among SUs and limiting the protocol overhead.
However, the achieved throughput is less sensitive to the transmission probability  $p$ as will be demonstrated later via the numerical study.
Therefore, we will seek to optimize the throughput over $P_{\sf sen}$ and $T_S$ for a reasonable and fixed value of $p$. 

 For brevity, we express the throughput as a function of $P_{\sf sen}$ and $T_S$ only, i.e., $\mathcal{NT}(T_S, P_{\sf sen})$.
Suppose that the PU requires that the average detection probability is at least $\overline{\mathcal{P}}_{d}$.
Then, the throughput maximization problem can be stated as follows:
\begin{equation}
\label{eq3a}
\begin{array}{l}
 {\mathop {\max }\limits_{T_S, P_{\sf sen}}} \quad {\mathcal{NT}} \left(T_S, P_{\sf sen}\right)  \\ 
 \mbox{s.t.}\,\,\,\, \hat{\mathcal{P}}_{d}\left(\varepsilon,T_S\right) \geq \mathcal{\overline P}_{d},  \\
 \quad \quad 0 \leq P_{\sf sen} \leq P_{\sf max}, \quad 0 \leq T_S \leq T,\\
 \end{array}\!\!
\end{equation}
where $P_{\sf max}$ is the maximum power for SUs, and $T_S$ is upper bounded by $T$.
In fact, the first constraint on $\hat{\mathcal{P}}_{d}\left(\varepsilon,T_S\right)$ implies that the spectrum sensing should be sufficiently reliable to protect the PU which can
be achieved  with sufficiently large sensing time $T_S$. 
Moreover, the SU's transmit
 power $P_{\sf sen}$ must be appropriately set to achieve good tradeoff between the network throughput and self-interference mitigation.

\subsection{Parameter Configuration for FDC--MAC Protocol}

To gain insights into the parameter configuration of the FDC--MAC protocol, we first study the optimization with respect to the sensing time $T_S$ for a given $P_{\sf sen}$.
For any value of $T_S$, we would need to set the sensing detection threshold $\varepsilon$ 
so that  the detection probability constraint
is met with equality, i.e., $\mathcal{\hat P}_d \left(\varepsilon,T_S\right) = \mathcal{\overline P}_d$ as in \cite{Liang08, Tan11}. 
Since the detection probability is smaller in \textbf{Case 3} (i.e., the PU changes from the idle to 
active status during the FD sensing stage of duration $T_S$) compared to that in \textbf{Case 1} and \textbf{Case 2} (i.e.,
the PU remains idle during the FD sensing stage) considered in the previous section, we only need to consider \textbf{Case 3}
to maintain the detection probability constraint. The average probability of detection for the FD sensing in \textbf{Case 3} can be expressed as
\beqn
\label{P_average}
\mathcal{\hat P}_d = \!  \int_{0}^{T_S}  \mathcal{P}_d^{01}(t) \! f_{\tau_{\sf id}}\!\!\left(t\left|0 \leq t \leq T_S\right.\right) dt,
\eeqn
where $t$ denotes the duration from the beginning of the FD sensing stage to the instant when the PU changes to the active state, and
$f_{\tau_{\sf id}}\left(t\left|\mathcal{A}\right.\right)$ is the pdf of $\tau_{\sf id}$ conditioned on event $\mathcal{A}$
capturing the condition $0 \leq t \leq T_S$, which is given as
\beqn
\label{cond_pdf_tau_id}
f_{\tau_{\sf id}}\left(t\left|\mathcal{A}\right.\right) = \frac{f_{\tau_{\sf id}}\left(t\right)}{\Pr\left\{\mathcal{A}\right\}} = \frac{\frac{1}{{\bar \tau}_{\sf id}} \exp(-\frac{t}{{\bar \tau}_{\sf id}})}{1-\exp(-\frac{T_S}{{\bar \tau}_{\sf id}})}.
\eeqn
Note that $\mathcal{P}_d^{01}(t)$ is derived in Appendix \ref{CAL_P_F_P_D} and $f_{\tau_{\sf id}} \left(t\right)$ is given in (\ref{pdf_tau_ac_id}). 

We consider the following single-variable optimization problem for a given $P_{\sf sen}$:
\begin{equation}
\label{NT_NonFrag_OPT_TS}
{\mathop {\max }_{0< T_S \leq T} } \quad {\mathcal{NT}} \! \left(T_S,  {P}_{\sf sen} \right). 
\end{equation}

We characterize the properties of function $\mathcal{NT}(T_S, P_{\sf sen})$ with respect to $T_S$ for a given  $P_{\sf sen}$ in the following theorem
whose proof is provided in Appendix~\ref{Prosp1}. For simplicity,
the throughput function is written as  $\mathcal{NT}(T_S)$.

\vspace{0.2cm}
\noindent
\textbf{Theorem 1:} The objective function ${{\mathcal{NT}} \!\!\left( { T_S } \right)}$ of (\ref{NT_NonFrag_OPT_TS}) satisfies
the following properties
\begin{enumerate}
\item 
$\mathop {\lim }\limits_{T_S  \to 0} \frac{\partial {\mathcal{NT}}}{\partial T_S } =  + \infty$,

\item 
\begin{enumerate}
\item For HDTx mode with $\forall P_{\sf sen}$ and FDTx mode with $P_{\sf sen} < \overline{P}_{\sf sen}$, we have $\mathop {\lim }\limits_{T_S  \to T}   \frac{\partial {\mathcal{NT}}}{\partial T_S }  < 0$,

\item For FDTx mode with $P_{\sf sen} > \overline{P}_{\sf sen}$, we have $\mathop {\lim }\limits_{T_S  \to T}   \frac{\partial {\mathcal{NT}}}{\partial T_S }  > 0$,
\end{enumerate}

\item $\frac{\partial^2 {\mathcal{NT}}}{\partial T_S^2} < 0$, $\forall T_S$,

\item The objective function ${{\mathcal{NT}} \!\!\left( T_S \right)}$ is bounded from above,
\end{enumerate}
where $\overline{P}_{\sf sen} =  N_0 \left[\left(1+\frac{P_{\sf dat}}{N_0+ \zeta P_{\sf dat}^\xi}\right)^2-1\right]$ is the
critical value of ${P}_{\sf sen}$ such that 
 $\mathop {\lim } \limits_{T_S  \to T} \frac{\partial \mathcal{NT}}{\partial T_S} = 0$.


We would like to discuss the properties stated in Theorem 1. 
For the HDTx mode with $\forall P_{\sf sen}$ and FDTx mode with low $P_{\sf sen}$, then properties 1, 2a, and 4 imply that there must be at least 
one $T_S$ in  $\left[0,T\right]$ that maximizes $\mathcal{NT}\left( T_S \right)$. The third property implies that  this maximum is indeed unique. 
Moreover, for the FDTx with high $P_{\sf sen}$, then properties 1, 2b, 3 and 4 imply that ${{\mathcal{NT}} \!\!\left( { T_S } \right)}$ increases in $\left[0,T\right]$.
Hence, the throughput ${{\mathcal{NT}} \!\!\left( { T_S } \right)}$ achieves its maximum with sensing time $T_S = T$.
We propose an algorithm to determine optimal $\left(T_S, P_{\sf sen}\right)$, which is summarized in Algorithm~\ref{OPT_Throughput_GMAC}.
Here, we can employ the bisection scheme and other numerical methods to determine the optimal value $T_S$ for a given $P_{\sf sen}$.

\begin{algorithm}[h]
\caption{\textsc{FDC-MAC Configuration Algorithm}}
\label{OPT_Throughput_GMAC}
\begin{algorithmic}[1]

\FOR {each considered value of $P_{\sf sen} \in [0,P_{\sf max}]$}

\STATE Find optimal $T_S$ for problem (\ref{NT_NonFrag_OPT_TS}) using the bisection method as ${\overline T}_S \left(P_{\sf sen}\right) = \mathop {\argmax} \limits_{0 \leq T_S \leq T} \mathcal{NT} \left(T, P_{\sf sen}\right)$.

\ENDFOR

\STATE The final solution $\left(T_S^*, P_{\sf sen}^*\right)$ is determined as $\left(T_S^*, P_{\sf sen}^*\right) = \mathop {\argmax} \limits_{P_{\sf sen}, {\overline T}_S \left(P_{\sf sen}\right)} \mathcal{NT} \left(T_S\left(P_{\sf sen}\right), P_{\sf sen}\right)$.

\end{algorithmic}
\end{algorithm}

\section{Numerical Results}
\label{Results}

For numerical studies, we set the key parameters for the FDC--MAC protocol as follows: mini-slot duration is $\sigma = 20 {\mu} s$; $PD = 1 {\mu} s$; $SIFS = 2\sigma$ ${\mu} s$; $DIFS = 10\sigma$ ${\mu} s$; $ACK = 20\sigma$ ${\mu} s$; $CTS = 20\sigma$ ${\mu} s$; $RTS = 20\sigma$ ${\mu} s$. Other parameters are chosen as follows unless stated otherwise: the
sampling frequency $f_s = 6$ MHz; bandwidth of PU's signal $6$ MHz; $\mathcal{\overline P}_d = 0.8$; $T = 15$ ms; $p = 0.0022$; the SNR of the PU 
signal at each SU $\gamma_P = \frac{P_p}{N_0} = -20$ dB; varying self-interference parameters $\zeta$ and $\xi$. Without loss of generality, the noise power is
normalized to one; hence, the SU transmit power $P_{\sf sen}$ becomes $P_{\sf sen} = SNR_s$; and we set $P_{\sf max} = 15$dB.

\begin{figure}[!t]
\centering
\includegraphics[width=80mm]{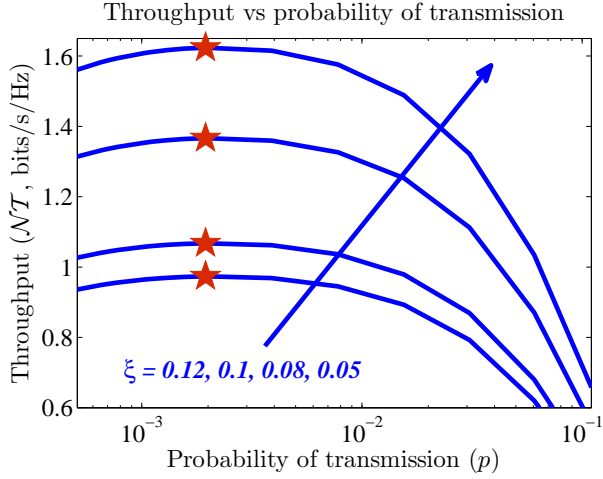}
\caption{Normalized throughput versus transmission probability $p$ for $T = 18$ ms, ${\bar \tau}_{\sf id} = 1000$ ms, ${\bar \tau}_{\sf ac} = 100$ ms, and varying $\xi$.}
\label{T_VS_p_nofrag_xi}
\end{figure}

\begin{figure}[!t]
\centering
\includegraphics[width=80mm]{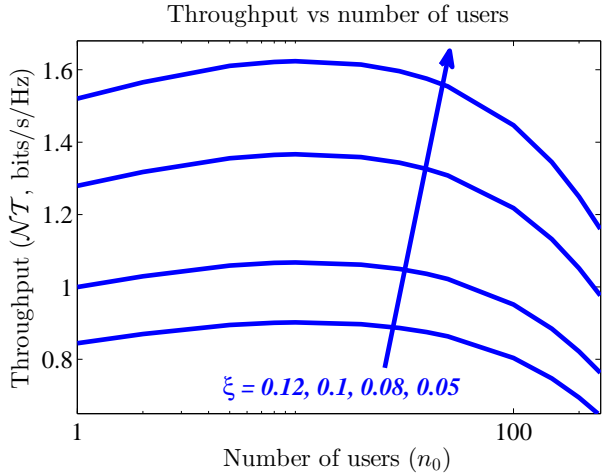}
\caption{Normalized throughput versus the number of SUs $n_0$ for $T = 18$ ms, $p = 0.0022$, ${\bar \tau}_{\sf id} = 1000$ ms, ${\bar \tau}_{\sf ac} = 100$ ms, and varying $\xi$.}
\label{T_VS_n0_xi_nofrag}
\end{figure}

We first study the impacts of self-interference parameters on the throughput performance with the following parameter setting: $\left({\bar \tau}_{\sf id}, {\bar \tau}_{\sf ac}\right) = 
\left(1000, 100\right)$ ms, $P_{\sf max} = 25$ dB, $T_{\sf eva} = 40$ ms, $\zeta = 0.4$, $\xi$ is varied in $\xi = \left\{0.12, 0.1, 0.08, 0.05\right\}$, and $P_{\sf dat} = P_{\sf max}$.
Recall that the self-interference depends on the transmit power $P$ as  $I(P) = \zeta \left(P\right)^{\xi}$ where $P= P_{\sf sen}$ and $P = P_{\sf dat}$
in the FD sensing and transmission stages, respectively. 
Fig.~\ref{T_VS_p_nofrag_xi} illustrates the variations of the throughput versus the transmission probability  $p$.
It can be observed that when $\xi$ decreases (i.e., the self-interference is smaller), the achieved throughput increases.
This is because SUs can transmit with higher power while still maintaining the sensing constraint during the FD sensing stage, which leads to throughput improvement. 
The optimal $P_{\sf sen}$ corresponding to these values of $\xi$ are $P_{\sf sen} = SNR_s = \left\{25.00, 18.01, 14.23, 11.28\right\}$ dB and 
the optimal probability of transmission is $p^* = 0.0022$ as indicated by a star symbol.
Therefore, to obtain all other results in this section, we set $p^* = 0.0022$.

Fig.~\ref{T_VS_n0_xi_nofrag} illustrates the throughput performance versus number of SUs $n_0$ when we keep the same parameter settings
as those for Fig.~\ref{T_VS_p_nofrag_xi} and $p^* = 0.0022$.
Again, when $\xi$ decreases (i.e., the self-interference becomes smaller), the achieved throughput increases. 
In this figure, the optimal $SNR_s$ achieving the maximum throughput corresponding to the considered values of $\xi$ 
are $P_{\sf sen} = SNR_s = \left\{25.00, 18.01, 14.23, 11.28\right\}$ dB, respectively.

\begin{figure}[!t]
\centering
\includegraphics[width=80mm]{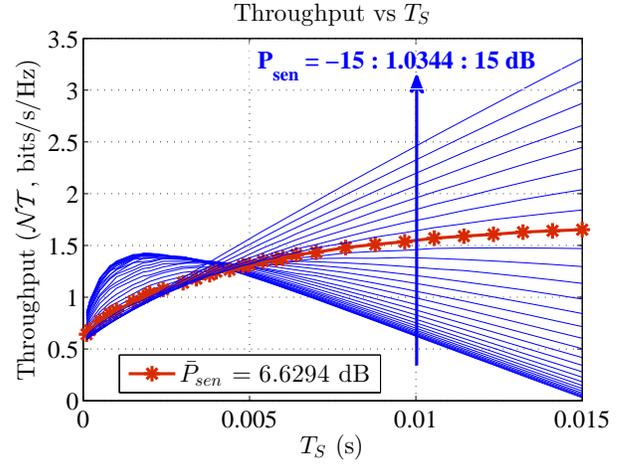}
\caption{Normalized throughput versus SU transmit power $P_{\sf sen}$ and sensing time $T_S$ for $p = 0.0022$, ${\bar \tau}_{\sf id} = 500$ ms, ${\bar \tau}_{\sf ac} = 50$ ms, $n_0 = 40$, 
$\xi =1$, $\zeta = 0.7$ and FDTx with $P_{\sf dat} = 15$ dB.}
\label{T_vs_Tsen_FDTX_Largezetaxi}
\end{figure}

We now verify the results stated in Theorem 1 for the FDTx mode. Specifically,
Fig.~\ref{T_vs_Tsen_FDTX_Largezetaxi} shows the throughput performance for the scenario where the QSIC is very low with large $\xi$ and $\zeta$
where we set the network parameters as follows: $p = 0.0022$, ${\bar \tau}_{\sf id} = 500$ ms, ${\bar \tau}_{\sf ac} = 50$ ms, $n_0 = 40$, $\xi =1$, $\zeta = 0.7$, and $P_{\sf dat} = 15$ dB.
Moreover, we can obtain $\overline{P}_{\sf sen}$ as in (\ref{P_sen_threshold}) in Appendix~\ref{Prosp1}, which is equal to $\overline{P}_{\sf sen} = 6.6294$ dB.
In this figure, the curve indicated by asterisks, which corresponds to $P_{\sf sen} = \overline{P}_{\sf sen}$, shows the monotonic increase of the throughput with sensing
time $T_S$ and other curves corresponding to $P_{\sf sen} > \overline{P}_{\sf sen}$ have the same characteristic.
In contrast, all remaining curves (corresponding to $P_{\sf sen} < \overline{P}_{\sf sen}$) first increase to the maximum values and then decrease as we increase $T_S$.

\begin{figure}[!t]
\centering
\includegraphics[width=80mm]{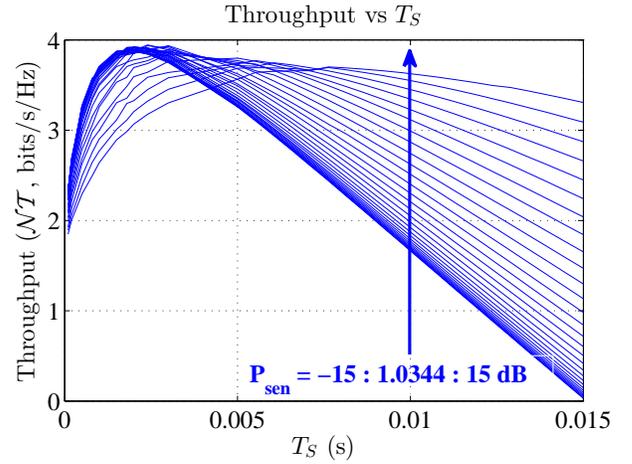}
\caption{Normalized throughput versus SU transmit power $P_{\sf sen}$ and sensing time $T_S$ for $p = 0.0022$, ${\bar \tau}_{\sf id} = 500$ ms, ${\bar \tau}_{\sf ac} = 50$ ms, $n_0 = 40$, $\xi =1$, $\zeta = 0.08$ and FDTx with $P_{\sf dat} = 15$ dB.}
\label{T_vs_Tsen_FDTX_Smallzetaxi}
\end{figure}

Fig.~\ref{T_vs_Tsen_FDTX_Smallzetaxi} illustrates the throughput performance for the very high QSIC with small $\xi$ and $\zeta$
where we set the network parameters as follows: $p = 0.0022$, ${\bar \tau}_{\sf id} = 500$ ms, ${\bar \tau}_{\sf ac} = 50$ ms, $n_0 = 40$, $\xi =1$, $\zeta = 0.08$, and $P_{\sf dat} = 15$ dB.
Moreover, we can obtain $\overline{P}_{\sf sen}$ as in (\ref{P_sen_threshold}) in Appendix~\ref{Prosp1}, which is equal to $\overline{P}_{\sf sen} = 19.9201$ dB.
We have $P_{\sf sen} < P_{\sf max} = 15dB < \overline{P}_{\sf sen}$ in this scenario; hence, all the curves first increases to the maximum throughput and then decreases with
the increasing $T_S$. Therefore, we have correctly validated the properties stated in Theorem 1.

\begin{figure}[!t]
\centering
\includegraphics[width=80mm]{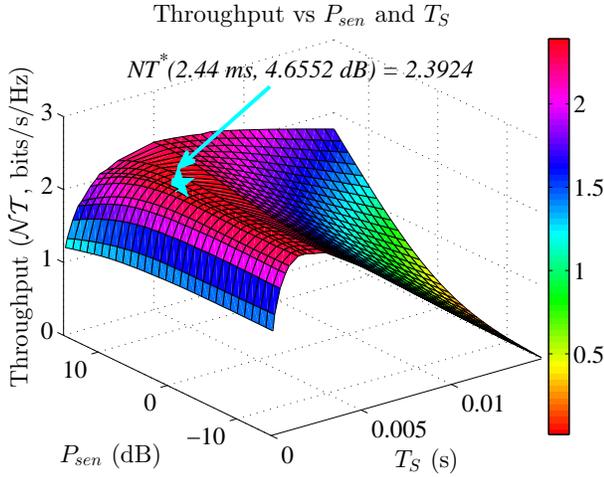}
\caption{Normalized throughput versus SU transmit power $P_{\sf sen}$ and sensing time $T_S$ for $p = 0.0022$, ${\bar \tau}_{\sf id} = 150$ ms, ${\bar \tau}_{\sf ac} = 50$ ms, 
$n_0 = 40$, $\xi =0.95$, $\zeta = 0.08$ and FDTx with $P_{\sf dat} = 15$ dB.}
\label{P2_15dB_tau1505015ms}
\end{figure}

Now we investigate the throughput performance versus SU transmit power $P_{\sf sen}$ and sensing time $T_S$ for the case of high QSIC with $\xi =0.95$ and $\zeta = 0.08$.
Fig.~\ref{P2_15dB_tau1505015ms} shows the throughput versus the SU transmit power $P_{\sf sen}$ and sensing time $T_S$ for the FDTx mode with $P_{\sf dat} = 15$ dB, $p = 0.0022$, 
${\bar \tau}_{\sf id} = 150$ ms, ${\bar \tau}_{\sf ac} = 50$ ms, and $n_0 = 40$. 
It can be observed that there exists an optimal configuration of the SU transmit power $P_{\sf sen}^* = 4.6552$ dB and sensing time $T_S^* = 2.44$ ms to achieve the 
maximum throughput $\mathcal{NT}\left(T_S^*, P_{\sf sen}^*\right) = 2.3924$, which is indicated by a star symbol.
These results confirm that SUs must set appropriate sensing time and transmit power for the FDC--MAC protocol to achieve the maximize throughput,
which cannot be achieved by setting $T_s = T$ as proposed in existing designs such as in \cite{report}. 

%

\begin{figure}[!t]
\centering
\includegraphics[width=80mm]{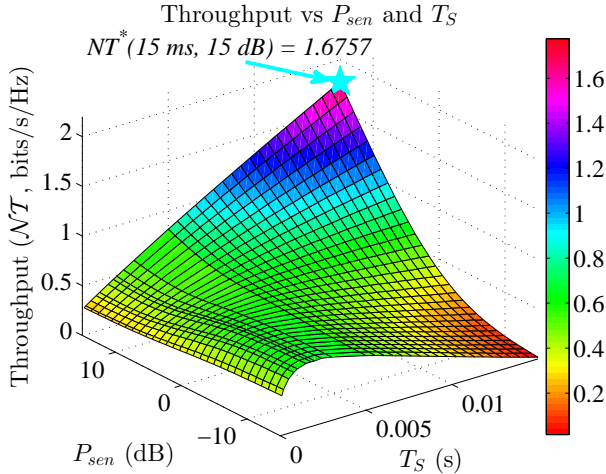}
\caption{Normalized throughput versus SU transmit power $P_{\sf sen}$ and sensing time $T_S$ for $p = 0.0022$, ${\bar \tau}_{\sf id} = 150$ ms, ${\bar \tau}_{\sf ac} = 50$ ms, $n_0 = 40$, $\xi =0.95$, $\zeta = 0.8$ and FDTx with $P_{\sf dat} = 15$ dB.}
\label{P2_15dB_tau1505015ms_Largezetaxi}
\end{figure}

In Fig.~\ref{P2_15dB_tau1505015ms_Largezetaxi}, we present the throughput versus the SU transmit power $P_{\sf sen}$ and sensing time $T_S$ 
for the low QSIC scenario where $p = 0.0022$, ${\bar \tau}_{\sf id} = 150$ ms, ${\bar \tau}_{\sf ac} = 50$ ms, $P_{\sf max} = 15$ dB, $n_0 = 40$, $\xi =0.95$, and $\zeta = 0.8$. 
The optimal configuration of SU transmit power $P_{\sf sen}^* = 15$ dB and sensing time $T_S^* = 15$ ms to achieve the maximum throughput $\mathcal{NT}\left(T_S^*, P_{\sf sen}^*\right) = 1.6757$
 is again indicated by a star symbol. Under this optimal configuration, the FD sensing is performed during the whole data phase (i.e., there is no transmission stage). 
In fact, to achieve the maximum throughput, the SU must provide the satisfactory sensing performance and attempt to achieve high transmission rate.
Therefore, if the QSIC is low, the data rate achieved during the transmission stage can be lower than that in the FD sensing stage because of
the very strong self-interference in the transmission stage. Therefore, setting longer FD sensing time enables to achieve more satisfactory sensing
performance and higher transmission rate, which explains that the optimal configuration should set $T_S^* = T$ for the low QSIC scenario.
This protocol configuration corresponds to existing design in \cite{report}, which is a special case of the proposed FDC--MAC protocol. 

\begin{figure}[!t]
\centering
\includegraphics[width=80mm]{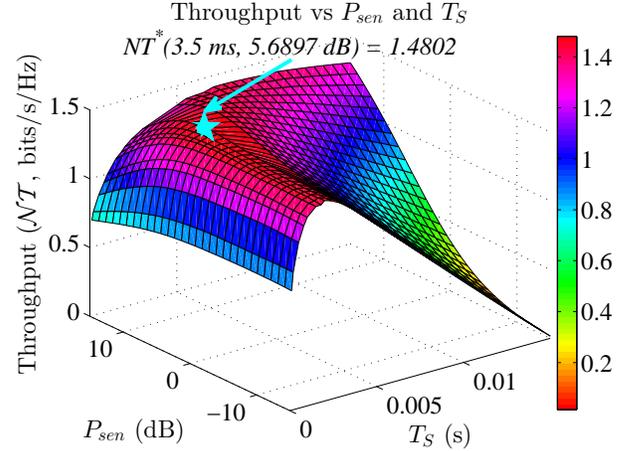}
\caption{Normalized throughput versus SU transmit power $P_{\sf sen}$ and sensing time $T_S$ for $p = 0.0022$, ${\bar \tau}_{\sf id} = 150$ ms, ${\bar \tau}_{\sf ac} = 50$ ms, $n_0 = 40$, $\xi =0.95$, $\zeta = 0.08$ and HDTx.}
\label{P2_15dB_tau1505015ms_HDTX}
\end{figure}

We now investigate the throughput performance with respect to the SU transmit power $P_{\sf sen}$ and sensing time $T_S$ for the HDTx mode.
Fig.~\ref{P2_15dB_tau1505015ms_HDTX} illustrates the throughput performance for the high QSIC scenario with $\xi =0.95$ and $\zeta = 0.08$.
It can be observed that there exists an optimal configuration of SU transmit power $P_{\sf sen}^* = 5.6897$ dB and sensing time $T_S^* = 3.5$ ms to achieve the maximum throughput $\mathcal{NT}\left(T_S^*, P_{\sf sen}^*\right) = 1.4802$, which is indicated by a star symbol.
The maximum achieved throughput of the HDTx mode is lower than that in the FDTx mode presented in Fig.~\ref{P2_15dB_tau1505015ms}.
This is because with high QSIC, the FDTx mode can transmit more data than the HDTx mode in the transmission stage.

\begin{figure}[!t]
\centering
\includegraphics[width=80mm]{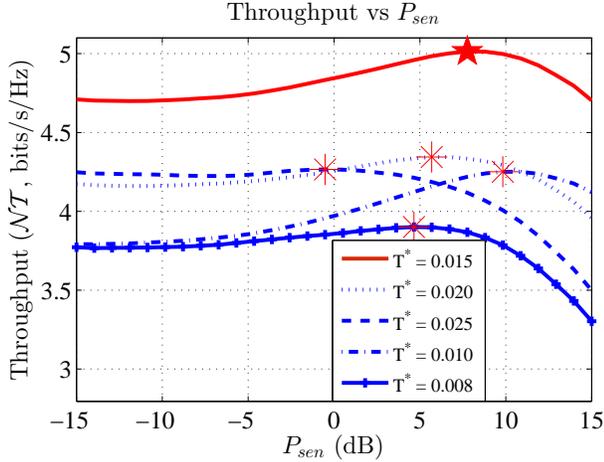}
\caption{Normalized throughput versus SU transmit power $P_{\sf sen}$ for $T_S = 2.2$ ms, $p = 0.0022$, ${\bar \tau}_{\sf id} = 1000$ ms, ${\bar \tau}_{\sf ac} = 50$ ms, $n_0 = 40$, 
$\xi =0.95$, $\zeta = 0.08$, varying $T$, and FDTx with $P_{\sf dat} = 15$ dB.}
\label{tauall_Tsens}
\end{figure}

In Fig.~\ref{tauall_Tsens}, we show the throughput versus the SU transmit power $P_{\sf sen}$ for $T_S = 2.2$ ms,  $p = 0.0022$, ${\bar \tau}_{\sf id} = 1000$ ms, ${\bar \tau}_{\sf ac} = 50$ ms,
 $n_0 = 40$, $\xi =0.95$, $\zeta = 0.08$ and various values of $T$ (i.e., the data phase duration) for the FDTx mode with $P_{\sf dat} = 15$ dB. 
For each value of $T$, there exists the optimal SU transmit power $P_{\sf sen}^*$ which is indicated by an asterisk.
It can be observed that as $T$ increases from 8 ms to 25 ms, the achieved maximum throughput first increases then decreases with $T$.
Also in the case with $T^* = 15$ ms, the SU achieves the largest throughput which is indicated by a star symbol.
Furthermore, the achieved throughput significantly decreases when the pair of $\left(T,P_{\sf sen}\right)$ deviates from the optimal values, $\left(T^*,P_{\sf sen}^*\right)$.

\begin{figure}[!t]
\centering
\includegraphics[width=80mm]{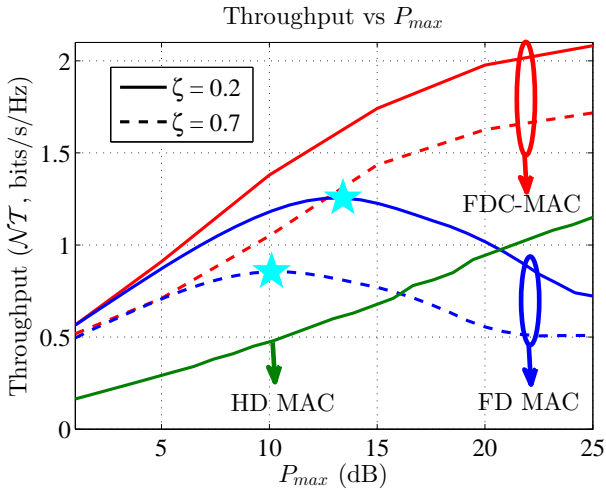}
\caption{Normalized throughput versus $P_{\sf max}$ for ${\bar \tau}_{\sf id} = 150$ ms, ${\bar \tau}_{\sf ac} = 75$ ms, $n_0 = 40$, $\xi =0.85$, $n_0 = 40$, 
$\zeta = \left\{0.2, 0.7\right\}$, and FDTx with $P_{\sf dat} = P_{\sf max}$ dB.}
\label{T_vs_Pmax_tauidac_150_75}
\end{figure}

Finally, we compare the throughput of our proposed FDC-MAC protocol,  the single-stage FD MAC protocol
where FD sensing (concurrent spectrum sensing and transmission)  is performed during the whole data phase \cite{report} and the HD MAC 
protocol which does not allow the transmission during the spectrum sensing interval in Fig.~\ref{T_vs_Pmax_tauidac_150_75}.
For brevity, the single-stage FD MAC protocol is refereed to as FD MAC in this figure.
The parameter settings are as follows: ${\bar \tau}_{\sf id} = 150$ ms, ${\bar \tau}_{\sf ac} = 75$ ms, $n_0 = 40$, $\xi =0.85$, $n_0 = 40$, 
$\zeta = \left\{0.2, 0.7\right\}$, and FDTx with $P_{\sf dat} = P_{\sf max}$ dB.
For fair comparison, we first obtain the optimal configuration of the single-stage FD MAC protocol, i.e., then we use $\left(T^*, p^*\right)$ for the HD MAC protocol and 
FDC-MAC protocol.
For the single-stage FD MAC protocol, the transmit power  is set to $P_{\sf max}$ because there is only a single stage where the SU performs 
sensing and transmission simultaneously during the data phase.
In addition, the HD MAC protocol will also transmit with the maximum transmit power $P_{\sf max}$ to achieve the highest throughput.
For both studied cases of $\zeta= \left\{0.2, 0.7\right\}$, our proposed FDC-MAC protocol significantly outperforms the other two protocols.
Moreover, the single-stage FD MAC protocol \cite{report} with power allocation outperforms the HD MAC protocol at the corresponding optimal value of $P_{\sf max}$
required by the single-stage FD MAC protocol. However, both single-stage FDC-MAC and HD MAC protocols achieve increasing throughput with higher $P_{\sf max}$ while
the single-stage FD MAC protocol has the throughput first increased then decreased as $P_{\sf max}$ increases. This demonstrates
the negative self-interference effect on the throughput performance of the single-stage FD MAC protocol, which 
is efficiently mitigated by our proposed FDC-MAC protocol.

\section{Conclusion}
\label{conclusion} 
In this paper, we have proposed the FDC--MAC protocol for cognitive radio networks, analyzed its throughput
performance, and studied its optimal parameter configuration. The design and analysis have taken into
account the FD communication capability and the self-interference of the FD transceiver.
We have shown that there exists an optimal FD sensing time to achieve the maximum throughput.
We have then presented extensive numerical results to demonstrate the impacts
of self-interference and protocol parameters on the throughput performance. In particular, we have shown that
the FDC--MAC protocol achieves significantly higher throughput than the HD MAC protocol, which confirms
that the FDC--MAC protocol can efficiently exploit the FD communication capability. Moreover,
the FDC--MAC protocol results in higher throughput with the increasing maximum power budget while
the throughput of the single-stage FD MAC can decrease in the high power regime. This result validates
the importance of adopting the two-stage procedure in the data phase and the optimization of sensing time
and transmit power during the FD sensing stage to mitigate the negative self-interference effect.

\appendices

\section{Derivation of ${\overline T}_{\sf cont}$}
\label{AppenA0}

To calculate ${\overline T}_{\sf cont}$, we define some further parameters as follows.
Denote $T_{\sf coll}$ as the duration of the collision and
$T_{\sf succ}$ as the required time for successful RTS/CTS transmission. These quantities can be calculated as follows \cite{Cali00}:
\beqn
\label{TCTSTI}
\left\{ {\begin{array}{*{20}{c}}
   T_{\sf succ} = DIFS + RTS + SIFS + CTS + 2PD \hfill  \\
   T_{\sf coll} = DIFS + RTS + PD, \hfill  \\
\end{array}} \right.
\eeqn
where $DIFS$ is the length of a DCF (distributed coordination function) interframe space, $RTS$ and $CTS$ denote
the lengths of the RTS and CTS messages, respectively.

As being shown in Fig.~\ref{Sentime_FDMAC_SF}, there can be several idle periods and collisions before one successful channel reservation.
Let $T_{\sf idle}^i$ denote the $i$-th idle duration between two consecutive RTS/CTS exchanges, which can be collisions or successful exchanges. 
Then, $T_{\sf idle}^i$ can be calculated based on its probability mass function (pmf),  which is derived as follows. 
In the following, all relevant quantities are defined in terms of the number of time slots.
With $n_0$ SUs joining the contention resolution, let $\mathcal{P}_{\sf succ}$, $\mathcal{P}_{\sf coll}$ and $\mathcal{P}_{\sf idle}$ denote
 the probabilities that a particular generic slot corresponds to a successful transmission, a collision, and an idle slot, respectively. These 
probabilities can be calculated as follows:
\beqn
\mathcal{P}_{\sf succ} &=& n_0p\left(1-p\right)^{n_0-1} \\
\mathcal{P}_{\sf idle} &=&  \left(1-p\right)^{n_0} \hspace{1cm} \\
\mathcal{P}_{\sf coll} &=&  1-\mathcal{P}_{\sf succ}-\mathcal{P}_{\sf idle},
\eeqn
where $p$ is the transmission probability of an SU in a generic slot. 
In general, the interval ${ T}_{\sf cont}$, whose average value is  ${\overline T}_{\sf cont}$ given in (\ref{tover}), consists of several intervals 
corresponding to idle periods, collisions, and one successful RTS/CTS transmission. 
Hence, this quantity can be expressed as 
\beqn
\label{T_cont}
{T}_{\sf cont} = \sum_{i=1}^{N_{\sf coll}} \left(T_{\sf coll}+ T_{\sf idle}^i\right) + T_{\sf idle}^{N_{\sf coll}+1} + T_{\sf succ},
\eeqn
where $N_{\sf coll}$ is the number of collisions before the successful RTS/CTS exchange and 
$N_{\sf coll}$ is a geometric random variable (RV) with parameter $1-\mathcal{P}_{\sf coll}/\mathcal{\overline P}_{\sf idle}$ where $\mathcal{\overline P}_{\sf idle} = 1 - \mathcal{P}_{\sf idle}$. 
Therefore, its pmf can be expressed as
\beqn
\label{N_c_cal}
 f_{X}^{N_{\sf coll}} \left(x\right) = \left(\frac{\mathcal{P}_{\sf coll}}{\mathcal{\overline P}_{\sf idle}}\right)^{x} \left(1-\frac{\mathcal{P}_{\sf coll}}{\mathcal{\overline P}_{\sf idle}}\right), \: x = 0, 1, 2, \ldots
\eeqn
Also, $T_{\sf idle}$ represents the number of consecutive idle slots, which is also a geometric RV with parameter $1-\mathcal{P}_{\sf idle}$ with the following pmf
\beqn
\label{T_I_cal}
f_{X}^{T_{\sf idle}} \left(x\right) = \left(\mathcal{P}_{\sf idle}\right)^{x} \left(1-\mathcal{P}_{\sf idle}\right), \: x = 0, 1, 2, \ldots
\eeqn
Therefore, ${\overline T}_{\sf cont}$ (the average value of ${T}_{\sf cont}$) can be written as follows \cite{Cali00}:
\beqn
{\overline T}_{\sf cont}  = {\overline N}_{\sf coll}T_{\sf coll} + {\overline T}_{\sf idle} \left({\overline N}_{\sf coll}+1\right) + T_{\sf succ} \label{T_contgeo},
\eeqn
where ${\overline T}_{\sf idle}$ and ${\overline N}_{\sf coll}$ can be calculated as
\beqn
{\overline T}_{\sf idle} &=& \frac{\left(1-p\right)^{n_0}}{1-\left(1-p\right)^{n_0}} \\
{\overline N}_{\sf coll} &=& \frac{1-\left(1-p\right)^{n_0}}{n_0p\left(1-p\right)^{n_0-1}}-1. 
\eeqn
These expressions are obtained by using the  pmfs of the corresponding RVs given in (\ref{N_c_cal}) and (\ref{T_I_cal}), respectively \cite{Cali00}.

\section{Derivations of $\mathcal{B}_1$, $\mathcal{B}_2$, $\mathcal{B}_3$}
\label{AppenB1}

We will employ a pair of parameters $\left(\theta, \varphi\right)$ to represent the HDTX and FDTX modes where ($\left(\theta, \varphi\right) = \left(0, 1\right)$) for
HDTx mode and  ($\left(\theta, \varphi\right) = \left(1, 2\right)$) for the FDTx mode. Moreover,
since the transmit powers in the FD sensing and transmission stages are different, which are equal to $P_{\sf sen}$ and $P_{\sf dat}$, respectively,
we define different SNRs and SINRs in these two stages as follows: $\gamma_{S1} = \frac{P_{\sf sen}}{N_0}$ and $\gamma_{S2} = \frac{P_{\sf sen}}{N_0+P_p}$
are the SNR and SINR achieved by the SU in the FD sensing stage with and without the presence of the PU, respectively; 
$\gamma_{D1} = \frac{P_{\sf dat}}{N_0+\theta I}$ and $\gamma_{D2} = \frac{P_{\sf dat}}{N_0+P_p+\theta I}$ for $I = \zeta P_{\sf dat}^\xi$ are the SNR and SINR
achieved by the SU in the transmission stage with and without the presence of the PU, respectively. It can be seen that we have accounted for
the self-interference for the FDTx mode during the transmission stage in $\gamma_{D1}$ by noting that $\theta=1$ in this case. The parameter $\varphi$
for the HDTx and FDTx modes will be employed to capture the throughput for one-way and two-way transmissions in these modes, respectively.

The derivations of $\mathcal{B}_1$, $\mathcal{B}_2$, and $\mathcal{B}_3$ require us to consider different possible
sensing outcomes in the FD sensing stage. In particular, we need to determine the detection probability $\mathcal{P}_d^{ij}$, which is the probability
of correctly detecting the PU given the PU is active, and the false alarm probability $\mathcal{P}_f^{ij}$, which is the
probability of the erroneous sensing of an idle channel, for each event $h_{ij}$ capturing the state changes of the PU. 
In the following analysis, we assume the exponential distribution for ${\tau}_{\sf ac}$ and ${\tau}_{\sf id}$ where ${\bar \tau}_{\sf ac}$ and ${\bar \tau}_{\sf id}$ denote the
corresponding average values of these active and idle intervals.
Specifically, let $f_{\tau_{\sf x}}\left(t\right)$ denote the pdf of $\tau_{\sf x}$ (${\sf x}$ represents ${\sf ac}$ or ${\sf id}$ in the pdf of $\tau_{\sf ac}$ or $\tau_{\sf id}$, respectively) 
then
\beqn
\label{pdf_tau_ac_id}
f_{\tau_{\sf x}}\left(t\right) =  \frac{1}{{\bar \tau}_{\sf x}} \exp(-\frac{t}{{\bar \tau}_{\sf x}}).
\eeqn
Similarly, we employ $T_S^{ij}$ and $T_D^{ij}$ to denote the number of bits transmitted on one unit of system bandwidth during the FD sensing and transmission stages under
the PU's state-changing event $h_{ij}$, respectively. 

We can now  calculate $\mathcal{B}_1$ as follows:
\beqn
\mathcal{B}_1 = \mathcal{P}\left(\mathcal{H}_0\right) \int_{t=T_{\sf ove}+T}^\infty T_1^{00} f_{\tau_{\sf id}}(t) dt \hspace{1cm} \nonumber\\
= \mathcal{P}\left(\mathcal{H}_0\right) T_1^{00} \exp\left(-\frac{T_{\sf ove}+T}{\bar{\tau}_{\sf id}}\right), \hspace{0.85cm}
\eeqn
where $\mathcal{P}\left(\mathcal{H}_0\right)$ denotes the probability of the idle state of the PU, and
  $\mathcal{P}_f^{00}$ is the false alarm probability for 
event $h_{00}$ given in Appendix~\ref{CAL_P_F_P_D}. Moreover, $T_1^{00} = \mathcal{P}_f^{00} T_S^{00} + (1-\mathcal{P}_f^{00})(T_S^{00}+T_D^{00})$, $T_S^{00} = T_S \log_2 \left(1+\gamma_{S1}\right)$, 
$T_D^{00} = \varphi \left(T-T_S\right) \log_2 \left(1+\gamma_{D1}\right)$ where $T_S^{00}$ and $T_D^{00}$ denote the number of bits transmitted (over one Hz of system bandwidth) in the FD sensing
and transmission stages of the data phase, respectively. 
After some manipulations, we achieve
\beqn 
\mathcal{B}_1 = \mathcal{K}_e  \exp \left(\frac{T}{\Delta\tau}\right) \left[T_S \log_2 \left(1+\gamma_{S1}\right) + \right. \hspace{0.4cm} \nonumber\\
\left.  \varphi \left(1 - \mathcal{P}_f^{00}\right) \left(T-T_S\right) \log_2 \left(1+\gamma_{D1}\right)\right], 
\eeqn
where $\mathcal{K}_e = \mathcal{P} \left(\mathcal{H}_0\right) \exp \left(-\left(\frac{T_{\sf ove}}{{\bar \tau}_{\sf id}}+\frac{T}{{\bar \tau}_{\sf ac}}\right)\right)$ and $\frac{1}{\Delta\tau} = \frac{1}{\bar{\tau}_{\sf ac}} - \frac{1}{\bar{\tau}_{\sf id}}$.

Moreover, we can calculate $\mathcal{B}_2$ as
\beqn \label{eqT2}
\mathcal{B}_2 = \mathcal{P}\left(\mathcal{H}_0\right) \int_{t_1 =T_{\sf ove}+T_S}^{T_{\sf ove}+T} \int_{t_2 =T_{\sf ove}+T-t_1}^\infty \hspace{2.4cm} \nonumber\\
T_2^{01}(t_1) f_{\tau_{\sf id}}(t_1)f_{\tau_{\sf ac}}(t_2) dt_1 dt_2,
\eeqn
where $T_2^{01}(t_1) = \mathcal{P}_f^{00} T_S^{00} + (1-\mathcal{P}_f^{00})(T_S^{00} +T_D^{01}\left(\bar{t}_1\right))$,
 $T_D^{01}\left({t}_1\right) = \varphi \left(T-T_S-\bar{t}_1\right) \log_2 \left(1+\gamma_{D2}\right) + \varphi \bar{t}_1 \log_2 \left(1+\gamma_{D1}\right)$, 
and $\bar{t}_1 = t_1 - \left(T_{\sf ove}+T_S\right)$. In this expression, $t_1$ denotes the interval from the beginning of the CA cycle to the instant when the PU changes to the active state from
an idle state. Again, $T_S^{00}$ and $T_D^{01}$ denote the amount of data transmitted in the FD sensing and transmission stages for this case, respectively.
 After some manipulations, we achieve
\beqn
\mathcal{B}_2 = \mathcal{K}_e \frac{\Delta\tau}{{\bar \tau}_{\sf id}} \left\{\left(\exp\left(\frac{T}{\Delta\tau}\right)-\exp\left(\frac{T_S}{\Delta\tau}\right)\right) \times \right. \nonumber \hspace{1.12cm}\\
 \left[T_S \log_2 \left(1\!+\!\gamma_{S1}\right) \!- \! \varphi \Delta\tau \left(1\!-\!\mathcal{P}_f^{00}\right) \log_2 \left(\frac{1\!+\!\gamma_{D1}}{1\!+\!\gamma_{D2}}\right)\right] \hspace{0.34cm} \nonumber \\
+ \varphi \left(T - T_S\right) \left(1\!-\!\mathcal{P}_f^{00}\right) \times \hspace{4.2cm} \nonumber\\
\left. \left[\!\exp\left(\!\frac{T}{\Delta\tau}\!\right) \log_2 \!\left(\!1\!+\!\gamma_{D1}\!\right) \!-\! \exp\left(\!\frac{T_S}{\Delta\tau}\!\right) \log_2 \left(1\!+\!\gamma_{D2}\right)\! \right]\! \right\}.
\eeqn

Finally, we can express $\mathcal{B}_3$ as follows:
\beqn
\mathcal{B}_3 = \mathcal{P}\left(\mathcal{H}_0\right) \int_{t_1 =T_{\sf ove}}^{T_{\sf ove}+T_S} \int_{t_2 =T_{\sf ove}+T-t_1}^\infty \nonumber \hspace{2.4cm}\\
\left[\mathcal{P}_d^{01} \left(\bar{t}_1\right) T_S^{01} \left(\bar{t}_1\right) + (1-\mathcal{P}_d^{01} \left(\bar{t}_1\right))(T_S^{01} \left(\bar{t}_1\right) + T_D^{11})\right] \nonumber \\
f_{\tau_{\sf id}}(t_1)f_{\tau_{\sf ac}}(t_2) dt_1 dt_2,
\eeqn
where $\bar{t}_1 = t_1 - T_{\sf ove}$, $T_S^{01} \left(\bar{t}_1\right) = \bar{t}_1 \log_2 \left(1+\gamma_{S1}\right) + \left(T_S-\bar{t}_1\right) \log_2 \left(1+\gamma_{S2}\right)$, $T_D^{11} = 
 \varphi \left(T-T_S\right) \log_2 \left(1+\gamma_{D2}\right)$, and $t_1$ is the same as in (\ref{eqT2}). Here, $T_S^{01}$ and $T_D^{11}$ denote the amount of data delivered
in the FD sensing and transmission stages for the underlying case, respectively. After some manipulations, we attain
\beqn
\mathcal{B}_3 = \mathcal{K}_e \int_{t =0}^{T_S} \left[T_S^{01} \left(t\right) + T_D^{11} -\mathcal{P}_d^{01} \left(t\right) T_D^{11}\right] \nonumber \\
f_{\tau_{\sf id}}(t) \exp\left(\frac{t}{\bar{\tau}_{\sf ac}}\right) dt = \mathcal{B}_{31} + \mathcal{B}_{32},
\eeqn
where
\beqn
\mathcal{B}_{31} = \mathcal{K}_e \int_{t =0}^{T_S} \left[T_S^{01} \left(t\right) + T_D^{11}\right] f_{\tau_{\sf id}}(t) \exp \left(\frac{t}{\bar{\tau}_{\sf ac}}\right) dt \nonumber \hspace{0.7cm}\\
= \mathcal{K}_e \frac{\Delta\tau}{\bar{\tau}_{\sf id}} \!\left\{\!\Delta\tau \!\left[\!\left(\!\frac{T_S}{\Delta\tau} \!-\! 1\!\right) \exp\! \left(\!\frac{T_S}{\Delta\tau}\!\right) \!+\! 1\right] \log_2 \!\left(\!\frac{1\!+\!\gamma_{S1}}{1\!+\!\gamma_{S2}}\!\right) \right. \nonumber \\
\left. + \left[\exp \left(\frac{T_S}{\Delta\tau}\right) - 1\right] \left[T_D^{11} + T_S \log_2 \left(1+\gamma_{S2}\right) \right]\right\}, \hspace{0.6cm}
\eeqn
and
\beqn \label{T32bar}
\mathcal{B}_{32} = -\mathcal{K}_e T_D^{11} \bar{T}_{32},
\eeqn
where $\bar{T}_{32} = \int_{t =0}^{T_S} \mathcal{P}_d^{01} \left(t\right) f_{\tau_{\sf id}}(t) \exp \left(\frac{t}{\bar{\tau}_{\sf ac}}\right) dt$.

\section{False Alarm and Detection Probabilities}
\label{CAL_P_F_P_D}

We derive the detection and false alarm probabilities for FD sensing and two PU's state-changing events $h_{00}$ and $h_{01}$ in this appendix.
Assume that the transmitted signals from the PU and SU are circularly symmetric complex Gaussian (CSCG) signals while the noise at the secondary
receiver is independently and identically distributed CSCG $\mathcal{CN}\left( {0,{N_0}} \right)$ \cite{Liang08}. 
Under FD sensing, the  false alarm probability for event $h_{00}$ can be derived using the similar method as in \cite{Liang08},
which is given as
\beqn
\mathcal{P}_f^{00} = \mathcal{Q} \left[\left(\frac{\epsilon}{N_0+I(P_{\sf sen})}-1\right)\sqrt{f_sT_S}\right],
\eeqn
where $\mathcal{Q} \left(x\right) = \int_x^{+\infty} \exp \left(-t^2/2\right) dt$; $f_s$, $N_0$, $\epsilon$, $I(P_{\sf sen})$ are the sampling frequency, the noise power, the detection
threshold and the self-interference, respectively; $T_S$ is the FD sensing duration. 


The  detection probability for event $h_{01}$ is given as
\beqn
\mathcal{P}_d^{01} \!\! =  \mathcal{Q} \left( \!\!\frac{\left(\!\! \frac{\epsilon}{N_0+I(P_{\sf sen})}- \frac{T_S-t}{T_S}\gamma_{PS}-1\right) 
 \!\sqrt{f_sT_S}}{\sqrt{\frac{T_S-t}{T_S}\left(\gamma_{PS}+1\right)^2+\frac{t}{T_S}}} \!\! \right), 
\eeqn
where $t$ is the interval from the beginning of the data phase to the instant when the PU changes its state,
 $\gamma_{PS} = \frac{P_p}{N_0+I(P_{\sf sen})}$ is the signal-to-interference-plus-noise ratio (SINR) of the PU's signal at the SU. 

\section{Proof of Theorem 1}
\label{Prosp1}

The first derivative of $\mathcal{NT}$ can be written as follows:
\beqn
\frac{\partial \mathcal{NT}}{\partial T_S} = \frac{1}{T_{\sf ove}+T} \sum_{i=1}^3 \frac{\partial \mathcal{B}_i}{\partial T_S}.
\eeqn
We derive the first derivative of $\mathcal{B}_i$ ($i = 1, 2, 3$) in the following. Toward this end, we will employ the approximation of $\exp \left(x\right) \approx 1 + x$, $x = \frac{T_x}{\tau_x}$, $T_x \in \left\{T, T_S, T-T_S\right\}$, $\tau_x \in \left\{{\bar \tau}_{\sf id}, {\bar \tau}_{\sf ac}, \Delta \tau\right\}$ where recall that $\frac{1}{\Delta\tau} = \frac{1}{\bar{\tau}_{\sf ac}} - \frac{1}{\bar{\tau}_{\sf id}}$. This approximation holds under the assumption that $T_x << \tau_x$ since we can omit all higher-power terms $x^n$ for 
$n > 1$ from the Maclaurin series expansion
of function $\exp \left(x\right)$. Using this approximation, we can express the first derivative of $\mathcal{B}_1$ as
\beqn
\frac{\partial \mathcal{B}_1}{\partial T_S} = \mathcal{K}_e \exp \left(\frac{T}{\Delta \tau}\right) \left\{\log_2 \left(1+\gamma_{S1}\right) \right. \hspace{2.5cm} \nonumber\\
\left. - \varphi\left[\left(T-T_S\right)\frac{\partial \mathcal{P}_f^{00}}{\partial T_S} + \left(1 - \mathcal{P}_f^{00}\right)\right]\log_2 \left(1+\gamma_{D1}\right) \right\},
\eeqn
where $\frac{\partial \mathcal{P}_f^{00}}{\partial T_S}$ is the first derivative of $\mathcal{P}_f^{00}$ whose derivation is given in Appendix \ref{APPROX_PF}.

Moreover, the first derivative of $\mathcal{B}_2$ can be written as
\beqn
\frac{\partial \mathcal{B}_2}{\partial T_S} = \mathcal{K}_e \frac{\Delta \tau}{{\bar \tau}_{\sf id}} \times \hspace{5.5cm} \nonumber\\
\left\{\!\left[\exp \left(\frac{T}{\Delta \tau}\right) \!- \!\left(1+\frac{T}{\Delta \tau}\right) \exp \left(\frac{T_S}{\Delta \tau}\right)\right] \log_2 \left(1+\gamma_{S1}\right) \right. \nonumber\\
- \varphi\frac{\partial \mathcal{P}_f^{00}}{\partial T_S} \!\left[\!\Delta \tau \!\left(\!\exp\! \left(\!\frac{T_S}{\Delta \tau}\!\right)\! - \!\exp \!\left(\!\frac{T}{\Delta \tau}\!\right)\!\right) \!\log_2\! \left(\!\frac{1\!+\!\gamma_{D1}}{1\!+\!\gamma_{D2}}\!\right) \!+\! \right. \nonumber\\
\left.\left(T\!\!-\!\!T_S\right) \!\!\left(\!\!\exp \!\left(\!\!\frac{T}{\Delta \tau}\!\!\right) \!\log_2 \!\left(\!1\!+\!\gamma_{D1}\!\right) \!-\! \exp \!\left(\!\!\frac{T_S}{\Delta \tau}\!\!\right) \!\log_2 \!\left(\!1\!+\!\gamma_{D2}\!\right) \!\!\right) \!\!\right] \nonumber\\
+ \varphi \left(1 - \mathcal{P}_f^{00}\right) \left[-\frac{T-T_S}{\Delta \tau} \exp \left(\frac{T_S}{\Delta \tau}\right) \log_2 \left(1+\gamma_{D2}\right) -  \right. \nonumber\\
\left(\exp \!\left(\!\frac{T}{\Delta \tau}\!\right) \log_2 \!\left(1\!+\!\gamma_{D1}\right) \!-\! \exp \left(\!\frac{T_S}{\Delta \tau}\!\right) \log_2 \!\left(\!1\!+\!\gamma_{D2}\!\right)\right) \nonumber\\
\left.\left.+ \exp \left(\frac{T_S}{\Delta \tau}\right) \log_2 \left(\frac{1+\gamma_{D1}}{1+\gamma_{D2}}\right) \right] \right\}.
\eeqn

Finally, the first derivative of $\mathcal{B}_3$ can be written as
\beqn
\frac{\partial \mathcal{B}_3}{\partial T_S} = \frac{\partial \mathcal{B}_{31}}{\partial T_S} + \frac{\partial \mathcal{B}_{32}}{\partial T_S},
\eeqn
where
\beqn
\frac{\partial \mathcal{B}_{31}}{\partial T_S} \!\!=\!\! \mathcal{K}_e \!\frac{\Delta \tau}{{\bar \tau}_{\sf id}}\!\! \left\{\!\Delta \tau \!\left[\!1\!+\!\left(\!\frac{T_S}{\Delta \tau}\!-\!1\!\right) \!\exp\! \left(\!\frac{T_S}{\Delta \tau}\!\right)\!\right] \!\log_2 \!\!\left(\!\frac{1\!+\!\gamma_{S1}}{1\!+\!\gamma_{S2}}\!\right) \right. \nonumber\\
\left.+\!\!\left(\!\exp \!\left(\!\!\frac{T_S}{\Delta \tau}\!\!\right)\!-\!1\!\right)\!\! \left[\!T_S \!\log_2 \!\left(\!1\!+\!\gamma_{S2}\!\right)\! +\!\varphi\!\left(\!T\!-\!T_S\!\right) \!\log_2 \!\left(\!1\!+\!\gamma_{D2}\!\right)\!\right]\!\right\}. 
\eeqn
To obtain the derivative for $\mathcal{B}_{32}$, we note that $1 \leq \exp \left(\frac{t}{\bar{\tau}_{\sf ac}}\right) \leq \exp \left(\frac{T_S}{\bar{\tau}_{\sf ac}}\right)$ for $\forall t \in \left[0, T_S\right]$. Moreover,
from the results in (\ref{P_average}) and (\ref{cond_pdf_tau_id}) and using the definition of $\bar{T}_{32}$ in (\ref{T32bar}), we have $\mathcal{\overline P}_d \left(1-\exp\left(\frac{-T_S}{{\bar \tau}_{\sf id}}\right)\right) \leq \bar{T}_{32} \leq \mathcal{\overline P}_d \left(1-\exp\left(\frac{-T_S}{{\bar \tau}_{\sf id}}\right)\right) \exp \left(\frac{T_S}{\bar{\tau}_{\sf ac}}\right)$.
Using these results, the first derivative of $\mathcal{B}_{32}$ can be expressed as
\beqn
\frac{\partial \mathcal{B}_{32}}{\partial T_S} = - \mathcal{K}_e \mathcal{\overline P}_d \varphi\frac{T-2T_S}{{\bar \tau}_{\sf id}} \log_2 \left(1+\gamma_{D2}\right).
\eeqn

Therefore, we have obtained the first derivative of  $\mathcal{NT}$ and we are ready to prove the first statement of Theorem 1.
Substitute $T_S = 0$ to the derived $\frac{\partial \mathcal{NT}}{\partial T_S}$ and use the approximation $\exp \left(x\right) \approx 1 + x$, 
 we yield the following result after some manipulations
\beqn
\mathop {\lim } \limits_{T_S \to 0} \frac{\partial \mathcal{NT}}{\partial T_S} = - K_0 K_1 \mathop {\lim } \limits_{T_S  \to 0} \frac{\partial \mathcal{P}_f^{00}}{\partial T_S},
\eeqn
where $K_0 = \frac{1}{T_{\sf ove}+T} \mathcal{K}_e$ and
\beqn
K_1 = \varphi \left[T \left(1+\frac{T}{\Delta \tau}\right) + \frac{T^2}{{\bar \tau}_{\sf id}}\right] \log_2 \left(1+\gamma_{D1}\right) \hspace{1.2cm} \nonumber\\
+ \varphi \frac{T\Delta \tau}{{\bar \tau}_{\sf id}} \log_2 \left(1+\gamma_{D2}\right).
\eeqn
It can be verified that $K_0 > 0$, $K_1 > 0$ and $\mathop {\lim } \limits_{T_S  \to 0} \frac{\partial \mathcal{P}_f^{00}}{\partial T_S} = - \infty$ by using the derivations in Appendix \ref{APPROX_PF};
 hence, we have $\mathop {\lim } \limits_{T_S  \to 0} \frac{\partial \mathcal{NT}}{\partial T_S} = +\infty >0$. This completes the proof of the first statement of the theorem.

We now present the proof for the second statement of the theorem.
Substitute $T_S = T$ to $\frac{\partial \mathcal{NT}}{\partial T_S}$ and utilize the approximation $\exp \left(x\right) \approx 1 + x$, we yield 
\beqn
\mathop {\lim } \limits_{T_S  \to T} \frac{\partial \mathcal{NT}}{\partial T_S} = \frac{1}{T_{\sf ove}+T} \sum_{i=1}^3 \frac{\partial \mathcal{B}_i}{\partial T_S} (T),
\eeqn
where we have
\beqn
\frac{\partial \mathcal{B}_1}{\partial T_S} (T) = \mathcal{K}_e \left(1+\frac{T}{\Delta \tau}\right) \times \hspace{3.9cm} \nonumber\\
\left[\log_2\left(1+\gamma_{S1}\right) - \varphi \left(1-\mathcal{P}_f^{00} (T)\right) \log_2 \left(1+\!\gamma_{D1}\right)\right] \label{deriT1}\\
\frac{\partial \mathcal{B}_2}{\partial T_S} (T) = - \mathcal{K}_e \frac{T}{{\bar \tau}_{\sf id}} \log_2 \left(1+\gamma_{S1}\right)  \hspace{3.2cm}\\
\frac{\partial \mathcal{B}_{31}}{\partial T_S} (T) = \mathcal{K}_e \! \frac{T}{{\bar \tau}_{\sf id}} \! \left[\log_2 \!\left(\!1\!+\!\gamma_{S1}\!\right)\left(\!1\!+\!\gamma_{S2}\!\right)\!-\! \varphi \! \log_2 \! \left(\!1\!+\!\gamma_{D2}\!\right)\right]  \\
\frac{\partial \mathcal{B}_{32}}{\partial T_S} (T) = \mathcal{K}_e \varphi \frac{T}{{\bar \tau}_{\sf id}} \mathcal{\overline P}_d \log_2 \left(1+\gamma_{D2}\right).  \hspace{2.6cm} \nonumber
\eeqn

Omit all high-power terms in the expansion of $\exp(x)$ (i.e., $x^n$ with $n>1$) where $x = \frac{T_x}{\tau_x}$, $T_x \in \left\{T, T_S, T-T_S\right\}$, $\tau_x \in \left\{{\bar \tau}_{\sf id}, {\bar \tau}_{\sf ac}, \Delta \tau\right\}$, we yield
\beqn \label{ThpT0}
\mathop {\lim } \limits_{T_S  \to T} \frac{\partial \mathcal{NT}}{\partial T_S} \approx \frac{1}{T_{\sf ove}+T}  \frac{\partial \mathcal{B}_1}{\partial T_S} (T). 
\eeqn

We consider the HDTx and FDTx modes in the following. For the HDTx mode, we have $\varphi = 1$ and $\theta = 0$. 
Then, it can be verified that $\mathop {\lim } \limits_{T_S  \to T} \frac{\partial \mathcal{NT}}{\partial T_S} < 0$ by using the results in (\ref{deriT1}) and (\ref{ThpT0}).
This is because we have $\log_2\left(1+\gamma_{S1}\right) - \left(1-\mathcal{P}_f^{00} (T)\right) \log_2 \left(1+\gamma_{D1}\right) \approx \log_2\left(1+\gamma_{S1}\right) - \log_2 \left(1+\gamma_{D1}\right) < 0$ (since
we have $\mathcal{P}_f^{00} (T) \approx 0$ and $\gamma_{S1} \leq \gamma_{D1}$).

For the FDTx mode, we have $\varphi = 2$, $\theta = 1$, and
also $\gamma_{S1} = \frac{P_{\sf sen}}{N_0}$ and $\gamma_{D1} = \frac{P_{\sf dat}}{N_0+I(P_{\sf dat})} = \frac{P_{\sf dat}}{N_0+ \zeta P_{\sf dat}^\xi}$.
We would like to define a critical value of ${P}_{\sf sen}$ which satisfies $\mathop {\lim } \limits_{T_S  \to T} \frac{\partial \mathcal{NT}}{\partial T_S} = 0$ 
to proceed further. Using the result in (\ref{deriT1}) and (\ref{ThpT0}) as well as the approximation $\mathcal{P}_f^{00} (T) \approx 0$, and
by solving $\mathop {\lim } \limits_{T_S  \to T} \frac{\partial \mathcal{NT}}{\partial T_S} = 0$ we yield
\beqn
\label{P_sen_threshold}
\overline{P}_{\sf sen} =  N_0 \left[\left(1+\frac{P_{\sf dat}}{N_0+\zeta P_{\sf dat}^\xi}\right)^2-1\right].
\eeqn
Using (\ref{deriT1}), it can be verified that if $P_{\sf sen} > \overline{P}_{\sf sen}$ then  $\mathop {\lim } \limits_{T_S  \to T} \frac{\partial \mathcal{NT}}{\partial T_S} > 0$;
otherwise, we have $\mathop {\lim } \limits_{T_S  \to T} \frac{\partial \mathcal{NT}}{\partial T_S} \leq 0$. So we have completed the proof for the second statement
of Theorem 1.

To prove the third statement of the theorem, we derive the second derivative of $\mathcal{NT}$ as
 \beq
\frac{\partial^2 \mathcal{NT}}{\partial T_S^2} = \frac{1}{T_{\sf ove}+T} 
\sum_{i=1}^3 \frac{\partial^2 \mathcal{B}_i}{\partial T_S^2},
\eeq
where we have 
\beqn \label{2derT1}
\frac{\partial^2 \mathcal{B}_1}{\partial T_S^2} = - \mathcal{K}_e\varphi \exp \left(\frac{T}{\Delta \tau}\right)  \log_2\left(1+\gamma_{D1}\right) \times \hspace{1.5cm} \nonumber\\
\left[\left(T-T_S\right) \frac{\partial^2 \mathcal{P}_f^{00}}{\partial T_S^2} - 2\frac{\partial \mathcal{P}_f^{00}}{\partial T_S}\right],
\eeqn
where $\frac{\partial^2 \mathcal{P}_f^{00}}{\partial T_S^2}$ is the second derivative of $\mathcal{P}_f^{00}$ and according to the derivations in Appendix~\ref{APPROX_PF}, we have 
$\frac{\partial^2 \mathcal{P}_f^{00}}{\partial T_S^2} > 0$, $\frac{\partial \mathcal{P}_f^{00}}{\partial T_S} < 0$, $\forall T_S$.
Therefore, we yield $\frac{\partial^2 \mathcal{B}_1}{\partial T_S^2} < 0$ $\forall T_S$.

Consequently, we have the following upper bound for $\frac{\partial^2 \mathcal{B}_1}{\partial T_S^2}$ by omitting the term $\exp \left(\frac{T}{\Delta \tau}\right) > 1$ in (\ref{2derT1})
\beqn
\frac{\partial^2 \mathcal{B}_1}{\partial T_S^2} \leq \mathcal{K}_e \left[h_1(T_S) + h_2(T_S) \right],
\eeqn
where 
\beqn
h_1(T_S) &=& - \varphi \left(T-T_S\right) \frac{\partial^2 \mathcal{P}_f^{00}}{\partial T_S^2} \log_2\left(1+\gamma_{D1}\right), \nonumber \\
h_2(T_S) &=& 2 \varphi \frac{\partial \mathcal{P}_f^{00}}{\partial T_S} \log_2\left(1+\gamma_{D1}\right). \nonumber
\eeqn
Moreover, we have
\beqn
\frac{\partial^2 \mathcal{B}_2}{\partial T_S^2} = \frac{\mathcal{K}_e \Delta \tau} {{\bar \tau}_{\sf id}} \left\{-\frac{2+\frac{T_S}{\Delta \tau}} {\Delta \tau} \exp \left(\frac{T_S}{\Delta \tau}\right)\log_2 \left(1+\gamma_{S1}\right)  \right. \hspace{0.2cm}\nonumber\\
-\!\varphi\frac{\partial^2 \mathcal{P}_f^{00}}{\partial T_S^2}\! \left[\!\Delta \tau \!\left(\!\exp\! \left(\!\frac{T_S}{\Delta \tau}\!\right)\! - \!\exp \!\left(\!\frac{T}{\Delta \tau}\!\right)\!\right) \!\log_2\! \left(\!\frac{1\!+\!\gamma_{D1}}{1\!+\!\gamma_{D2}}\!\right) \!\!+\!\!\right. \nonumber\\
\left.\left(\!T\!\!-\!\!T_S\!\right) \!\!\left(\!\!\exp \!\left(\!\!\frac{T}{\Delta \tau}\!\!\right) \!\log_2 \!\left(\!1\!+\!\gamma_{D1}\!\right) \!-\! \exp \!\left(\!\!\frac{T_S}{\Delta \tau}\!\!\right) \!\log_2 \!\left(\!1\!+\!\gamma_{D2}\!\right) \!\!\right) \!\!\right] \nonumber\\
+ 2\varphi\frac{\partial \mathcal{P}_f^{00}}{\partial T_S} \left[\left(\!\exp\! \left(\!\frac{T}{\Delta \tau}\!\right)\! - \!\exp \left(\!\frac{T_S}{\Delta \tau}\!\right)\!\right)\log_2 \!\left(1\!+\!\gamma_{D1}\right) \right. \hspace{0.6cm} \nonumber\\
\left.+\frac{T\!\!-\!\!T_S}{\Delta \tau} \exp \!\left(\!\!\frac{T_S}{\Delta \tau}\!\!\right) \log_2 \!\left(1\!+\!\gamma_{D2}\right)\right] \hspace{3.3cm} \nonumber\\
-\varphi\left(1 - \mathcal{P}_f^{00}\right) \frac{T\!\!-\!\!T_S}{\Delta \tau} \exp \!\left(\!\!\frac{T_S}{\Delta \tau}\!\!\right) \log_2 \!\left(1\!+\!\gamma_{D2}\right) \hspace{1.7cm} \nonumber\\
\left.+\varphi\left(1 - \mathcal{P}_f^{00}\right) \frac{1}{\Delta \tau} \exp \!\left(\!\!\frac{T_S}{\Delta \tau}\!\!\right) \log_2\! \left(\!\frac{1+\gamma_{D1}}{1+\gamma_{D2}}\!\right)\right\}. \hspace{1.2cm}
\eeqn
Therefore, we can approximate $\frac{\partial^2 \mathcal{B}_2}{\partial T_S^2}$ as follows:
\beqn
\frac{\partial^2 \mathcal{B}_2}{\partial T_S^2} = \mathcal{K}_e \left[h_3(T_S) + h_4(T_S) + h_5(T_S)\right],
\eeqn
where 
\beqn
h_3(T_S) = -\frac{\left(2+\frac{T_S}{\Delta \tau}\right)\left(1+\frac{T_S}{\Delta \tau}\right)}{{\bar \tau}_{\sf id}}  \log_2 \left(1+\gamma_{S1}\right), \hspace{1.10cm} \nonumber \\
h_4(T_S) = -\varphi\left(1 - \mathcal{P}_f^{00}\right) \frac{T\!\!-\!\!T_S}{{\bar \tau}_{\sf id}} \!\left(1+\!\frac{T_S}{\Delta \tau}\!\!\right) \log_2 \!\left(1\!+\!\gamma_{D2}\right) \hspace{0cm} \nonumber \\
 - \varphi\frac{\partial^2 \mathcal{P}_f^{00}}{\partial T_S^2} \left(\!T\! - \!T_S\!\right) \left[\!\frac{T}{{\bar \tau}_{\sf id}} \!\log_2\! \left(\!1+\!\gamma_{D1}\!\right)\! -\! \frac{T_S}{{\bar \tau}_{\sf id}} \!\log_2\! \left(\!1+\!\gamma_{D2}\!\right) \right] \hspace{0cm} \nonumber \\
  + 2\varphi\frac{\partial \mathcal{P}_f^{00}}{\partial T_S} \frac{T-T_S}{{\bar \tau}_{\sf id}} \!\!\left[\log_2\! \left(\!\frac{1\!+\!\gamma_{D1}}{1\!+\!\gamma_{D2}}\!\right) \!+\! \frac{T_S}{\Delta \tau} \log_2 \!\left(\!1\!+\!\gamma_{D2}\!\right) \right], \nonumber \\
h_5(T_S) = \varphi\left(1 - \mathcal{P}_f^{00}\right) \frac{1}{{\bar \tau}_{\sf id}} \!\left(1+\!\frac{T_S}{\Delta \tau}\!\!\right) \log_2\! \left(\!\frac{1+\gamma_{D1}}{1+\gamma_{D2}}\!\right). \hspace{0.34cm}\nonumber 
\eeqn
In addition, we have 
\beqn
\frac{\partial^2 \mathcal{B}_{31}}{\partial T_S^2} =  \frac{\mathcal{K}_e}{{\bar \tau}_{\sf id}} \exp \!\left(\!\!\frac{T_S}{\Delta \tau}\!\!\right) \left\{\left(1 + \frac{T}{\Delta \tau}\right)  \log_2 \left(1+\gamma_{S1}\right) \right. \hspace{0.1cm} \nonumber\\
\left.+ \log_2 \left(1+\gamma_{S2}\right) + \varphi\left(\!\frac{T-T_S}{\Delta \tau}\!-2\right)\log_2 \left(1+\gamma_{D2}\right)\right\}.
\eeqn
We can approximate $\frac{\partial^2 \mathcal{B}_{31}}{\partial T_S^2}$ as follows:
\beqn
\frac{\partial^2 \mathcal{B}_{31}}{\partial T_S^2} = \mathcal{K}_e \left[h_6(T_S) + h_6(T_S)\right],
\eeqn
where
\beqn
h_6(T_S) &=& \frac{1}{{\bar \tau}_{\sf id}} \log_2 \left(1\!+\!\gamma_{S1}\right) \left(1\!+\!\gamma_{S2}\right), \nonumber \\
h_7(T_S) &=& - \frac{2\varphi}{{\bar \tau}_{\sf id}}\log_2 \left(1\!+\!\gamma_{D2}\right).
\eeqn
Finally, we have
\beqn
\frac{\partial^2 \mathcal{B}_{32}}{\partial T_S^2} = \mathcal{K}_e h_8(T_S),
\eeqn
where 
\beqn
h_8(T_S) = \frac{2 \varphi \mathcal{\bar P}_d}{{\bar \tau}_{\sf id}} \log_2 \left(1\!+\!\gamma_{D2}\right).
\eeqn
The above analysis yields $\frac{\partial^2 \mathcal{NT}}{\partial T_S^2} = \mathcal{K}_e \sum_{i=1}^8 h_i(T_S)$. 
Therefore, to prove that $\frac{\partial^2 \mathcal{NT}}{\partial T_S^2} < 0$, we should prove that $h(T_S) < 0$ since $\mathcal{K}_e >0$ where
\beqn
h(T_S) = \sum_{i=1}^8 h_i(T_S).
\eeqn
It can be verified that $h_1(T_S) < 0$ and $h_4(T_S) < 0$, $\forall T_S$ because $\frac{\partial^2 \mathcal{P}_f^{00}}{\partial T_S^2} > 0$, $\frac{\partial \mathcal{P}_f^{00}}{\partial T_S} < 0$
according to Appendix \ref{APPROX_PF} and $\gamma_{D2} < \gamma_{D1}$. Moreover, we have
\beqn
h_3(T_S) < -\frac{2}{{\bar \tau}_{\sf id}}  \log_2 \left(1+\gamma_{S1}\right),
\eeqn
and because $\gamma_{S1} > \gamma_{S2}$, we have
\beqn
h_3(T_S) < -\frac{1}{{\bar \tau}_{\sf id}}  \log_2 \left(1+\gamma_{S1}\right)\left(1+\gamma_{S2}\right) = -h_6(T_S).
\eeqn
Therefore, we have $h_3(T_S) + h_6(T_S) < 0$.
Furthermore, we can also obtain the following result $h_7(T_S) + h_8(T_S) \leq 0$ because $\mathcal{\bar P}_d \leq 1$.
To complete the proof, we must prove that $h_2 (T_S) + h_5 (T_S) \leq 0$, which is equivalent to
\beqn
\label{EQN_LAST}
-2 {\bar \tau}_{\sf id} \frac{\partial \mathcal{P}_f^{00}}{\partial T_S} \frac{\log_2\left(1+\gamma_{D1}\right)}{\log_2\! \left(\!\frac{1+\gamma_{D1}}{1+\gamma_{D2}}\!\right)}  \geq \left(1 - \mathcal{P}_f^{00}\right)  \!\left(1+\!\frac{T_S}{\Delta \tau}\!\!\right),
\eeqn
where according to Appendix \ref{APPROX_PF}
\beqn
\frac{\partial \mathcal{P}_f^{00}}{\partial T_S} = - \frac{\bar{\gamma}\sqrt{f_s T_S}}{2\sqrt{2\pi}T_S}  \exp \left(-\frac{\left(\bar{\alpha}+\bar{\gamma} \sqrt{f_s T_S}\right)^2}{2} \right),
\eeqn
where $\bar{\alpha} = \left(\bar{\gamma}_1+1\right)\mathcal{Q}^{-1}\left(\overline{\mathcal{P}}_d\right)$. 
It can be verified that (\ref{EQN_LAST}) indeed holds because the LHS of (\ref{EQN_LAST}) is always larger than to 2 while the RHS of (\ref{EQN_LAST}) is always less than 2. 
Hence, we have completed the proof of the third statement of Theorem 1.

Finally,  the  fourth statement in the theorem obviously holds because $\mathcal{B}_i$ ($i$ = 1, 2, 3) are all bounded from above.
Hence, we have completed the proof of Theorem 1.

\section{Approximation of $\mathcal{P}_f^{00}$ and Its First and Second Derivatives}
\label{APPROX_PF}

We can approximate $\mathcal{\hat P}_d$ in (\ref{P_average}) as follows:
\beqn
\mathcal{\hat P}_d = \mathcal{Q} \left[\left(\frac{\epsilon}{N_0+I}-\bar{\gamma}-1\right)\frac{\sqrt{f_sT_S}}{\bar{\gamma}_1+1}\right],
\eeqn
where $\bar{\gamma}$ and $\bar{\gamma}_1$ are evaluated by a numerical method.
Hence, $\mathcal{P}_f^{00}$ can be calculated as we set $\mathcal{\hat P}_d = \overline{\mathcal{P}}_d$, which is given as follows:
\beqn
\mathcal{P}_f^{00} = \mathcal{Q} \left(\bar{\alpha}+\bar{\gamma}\sqrt{f_sT_S}\right),
\eeqn
where $\bar{\alpha} = \left(\bar{\gamma}_1+1\right)\mathcal{Q}^{-1}\left(\overline{\mathcal{P}}_d\right)$.

We now derive the first derivative of $\mathcal{P}_f^{00}$ as
\beqn
\frac{\partial \mathcal{P}_f^{00}}{\partial T_S} = - \frac{\bar{\gamma}\sqrt{f_s T_S}}{2\sqrt{2\pi}T_S}  \exp \left(-\frac{\left(\bar{\alpha}+\bar{\gamma} \sqrt{f_s T_S}\right)^2}{2} \right).
\eeqn
It can be seen that $\frac{\partial \mathcal{P}_f^{00}}{\partial T_S} <0$ since $\bar{\gamma} > 0$.
Moreover, the second derivative of $\mathcal{P}_f^{00}$ is
\beqn
\frac{\partial^2 \mathcal{P}_f^{00}}{\partial T_S^2} = \frac{\bar{\gamma}\sqrt{f_s T_S}}{4\sqrt{2\pi}T_S^2}  \left(1+\frac{1}{2}y\bar{\gamma} \sqrt{f_s T_S} \right)\exp \left(-\frac{y^2}{2} \right),
\eeqn
where $y=\bar{\alpha}+\bar{\gamma} \sqrt{f_s T_S}$.

We can prove that $\frac{\partial^2 \mathcal{P}_f^{00}}{\partial T_S^2} > 0$ by considering two different cases as follows. 
For the first case with $\frac{\bar{\alpha}^2}{\bar{\gamma}^2f_s} \leq T_S \leq T$ ($0 \leq \mathcal{P}_f^{00} \leq 0.5$), this statement holds since $y > 0$.
For the second case with $0 \leq T_S \leq \frac{\bar{\alpha}^2}{\bar{\gamma}^2f_s}$ ($0.5 \leq \mathcal{P}_f^{00} \leq 1$), $y \leq 0$, then we have
 $0< y-\bar{\alpha} = \bar{\gamma} \sqrt{f_s T_S} \leq -\bar{\alpha}$ and $0  \leq -y \leq -\bar{\alpha}$.
By applying the Cauchy-Schwarz inequality, we obtain $0 \leq -y (y-\bar{\alpha}) \leq \frac{\bar{\gamma}^2}{4} < 1 <2$; hence $1+\frac{1}{2}y\bar{\gamma} \sqrt{f_s T_S} > 0$. 
This result implies that $\frac{\partial^2 \mathcal{P}_f^{00}}{\partial T_S^2} > 0$.

\bibliographystyle{IEEEtran}


\begin{biography}[{\includegraphics[width=1in,height=1.25in,clip,keepaspectratio]{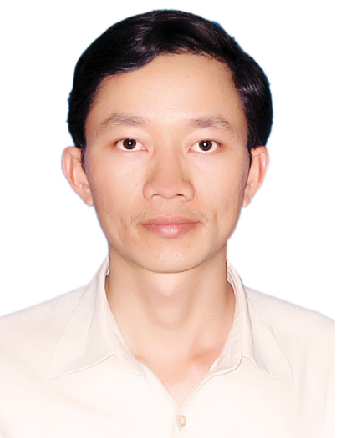}}]
{Le Thanh Tan} (S'11--M'15)  
 received the B.Eng. and M.Eng. degrees from Ho Chi Minh City University of Technology in 2002 and 2004, respectively and the Ph.D. degree from Institut National de la Recherche Scientifique--\'{E}nergie, Mat\'{e}riaux et T\'{e}l\'{e}communications (INRS--EMT), Canada in 2015. He is currently a Postdoctoral Research Associate at \'{E}cole Polytechnique de Montr\'{e}al, Canada. Before that he worked as a lecturer at Ho Chi Minh City University of Technical Education from 2002 to 2010. His current research activities focus on internet of things (IOT over LTE/LTE--A network, cyber--physical systems, big data, distributed sensing and control), time series analysis and dynamic factor models (stationary and non--stationary), wireless communications and networking,  Cloud--RAN, cognitive radios (software defined radio architectures, protocol design, spectrum sensing, detection, and estimation), statistical signal processing, random matrix theory, compressed sensing, and compressed sampling. He has served on TPCs of different international conferences including IEEE CROWNCOM, VTC, PIMRC, etc. He is a Member of the IEEE.
\end{biography}

\begin{biography}[{\includegraphics[width=1in,height=1.25in,clip,keepaspectratio]{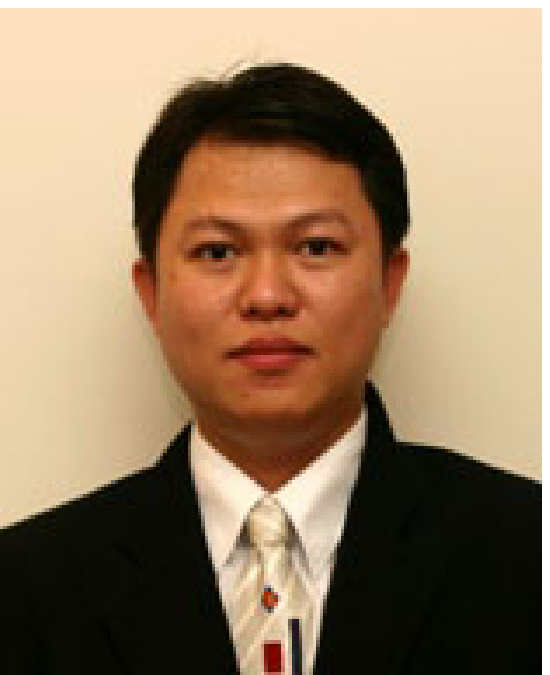}}]
{Long Le} (S'04--M'07--SM'12)  
received  the B.Eng.  degree  in  Electrical  Engineering  from  Ho Chi  Minh  City  University  of  Technology,  Vietnam, in 1999, the M.Eng. degree in Telecommunications from  Asian  Institute  of  Technology,  Thailand,  in 2002, and the Ph.D. degree in Electrical Engineering from  the  University  of  Manitoba,  Canada,  in  2007. He was a Postdoctoral Researcher at Massachusetts Institute  of  Technology  (2008--2010)  and  University  of  Waterloo  (2007--2008).  Since  2010,  he  has been  with  the  Institut  National  de  la  Recherche Scientifique  (INRS),  Universit\'{e} du Qu\'{e}bec, Montr\'{e}al,  QC,  Canada  where he  is  currently  an  associate  professor.  His  current  research  interests  include smart grids, cognitive radio, radio resource management, network control and optimization, and emerging enabling technologies for 5G wireless systems. He is a co-author of the book Radio Resource Management in Multi-Tier Cellular Wireless Networks (Wiley, 2013). Dr. Le is a member of the editorial board of  IEEE  TRANSACTIONS  ON  WIRELESS  COMMUNICATIONS,  IEEE COMMUNICATIONS SURVEYS AND TUTORIALS, and IEEE WIRELESS COMMUNICATIONS LETTERS. He has served as a technical program committee  chair/co-chair  for  several  IEEE  conferences  including  IEEE  WCNC, IEEE VTC, and IEEE PIMRC.
\end{biography}

\end{document}